\documentclass[fleqn,usenatbib]{mnras}

\usepackage{newtxtext,newtxmath}
\usepackage[T1]{fontenc}
\usepackage{ae,aecompl}
\usepackage{subfigure}
\usepackage{mathtools}

\usepackage{graphicx}	% Including figure files
\usepackage{amsmath}	% Advanced maths commands
\usepackage{amssymb}	% Extra maths symbols
\usepackage[dvipsnames]{xcolor}
\usepackage{hyperref}
\newcommand{\etal}{\textit{et al. }}

\def\CIVdblt{{\rm C~}\kern 0.1em{\sc iv}~$\lambda\lambda 1548, 1550$}
\def\NVdblt{{\rm N~}\kern 0.1em{\sc v}~$\lambda\lambda 1238, 1242$}
\def\OVIdblt{{\rm O~}\kern 0.1em{\sc vi}~$ 1031, 1037$}
\def\SiIVdblt{{\rm Si~}\kern 0.1em{\sc iv}~$\lambda\lambda1394, 1403$}
\def\NeVIII{\hbox{{\rm Ne~}\kern 0.1em{\sc viii}}}
\def\OII{\hbox{{\rm O~}\kern 0.1em{\sc ii}}}
\def\OIII{\hbox{{\rm O~}\kern 0.1em{\sc iii}}}
\def\OIV{\hbox{{\rm O~}\kern 0.1em{\sc iv}}}
\def\OV{\hbox{{\rm O~}\kern 0.1em{\sc v}}}
\def\OVI{\hbox{{\rm O~}\kern 0.1em{\sc vi}}}
\def\OVII{\hbox{{\rm O~}\kern 0.1em{\sc vii}}}
\def\OVIII{\hbox{{\rm O~}\kern 0.1em{\sc viii}}}
\def\NII{\hbox{{\rm N~}\kern 0.1em{\sc ii}}}
\def\NIII{\hbox{{\rm N~}\kern 0.1em{\sc iii}}}
\def\NIV{\hbox{{\rm N~}\kern 0.1em{\sc iv}}}
\def\NVII{\hbox{{\rm N~}\kern 0.1em{\sc vii}}}
\def\CII{\hbox{{\rm C~}\kern 0.1em{\sc ii}}}
\def\CIII{\hbox{{\rm C~}\kern 0.1em{\sc iii}}}
\def\SiIII{\hbox{{\rm Si~}\kern 0.1em{\sc iii}}}
\def\SIV{\hbox{{\rm S~}\kern 0.1em{\sc iv}}}
\def\SV{\hbox{{\rm S~}\kern 0.1em{\sc v}}}
\def\SVI{\hbox{{\rm S~}\kern 0.1em{\sc vi}}}
\def\SiII{\hbox{{\rm Si~}\kern 0.1em{\sc ii}}}
\def\SiIV{\hbox{{\rm Si~}\kern 0.1em{\sc iv}}}
\def\SiI{\hbox{{\rm Si~}\kern 0.1em{\sc i}}}
\def\PII{\hbox{{\rm P~}\kern 0.1em{\sc ii}}}
\def\AlII{\hbox{{\rm Al~}\kern 0.1em{\sc ii}}}
\def\AlIII{\hbox{{\rm Al~}\kern 0.1em{\sc iii}}}
\def\MgII{\hbox{{\rm Mg~}\kern 0.1em{\sc ii}}}
\def\FeII{\hbox{{\rm Fe~}\kern 0.1em{\sc ii}}}
\def\CaI{\hbox{{\rm Ca~}\kern 0.1em{\sc i}}}
\def\CaII{\hbox{{\rm Ca~}\kern 0.1em{\sc ii}}}
\def\CrII{\hbox{{\rm Cr~}\kern 0.1em{\sc ii}}}
\def\CII{\hbox{{\rm C~}\kern 0.1em{\sc ii}}}
\def\CIII{\hbox{{\rm C~}\kern 0.1em{\sc iii}}}
\def\CIV{\hbox{{\rm C~}\kern 0.1em{\sc iv}}}
\def\CV{\hbox{{\rm C}\kern 0.1em{\sc v}}}
\def\H{\hbox{{\rm H~}}}
\def\HI{\hbox{{\rm H~}\kern 0.1em{\sc i}}}
\def\HII{\hbox{{\rm H~}\kern 0.1em{\sc ii}}}
\def\Lya{\hbox{{\rm Ly}\kern 0.1em$\alpha$}}
\def\Lyb{\hbox{{\rm Ly}\kern 0.1em$\beta$}}
\def\Lyg{\hbox{{\rm Ly}\kern 0.1em$\gamma$}}
\def\Lyth{\hbox{{\rm Ly}\kern 0.1em$\theta$}}
\def\Lyfive{\hbox{{\rm Ly}\kern 0.1em$5$}}
\def\Lysix{\hbox{{\rm Ly}\kern 0.1em$6$}}
\def\Lyseven{\hbox{{\rm Ly}\kern 0.1em$7$}}
\def\Lyeight{\hbox{{\rm Ly}\kern 0.1em$8$}}
\def\Lynine{\hbox{{\rm Ly}\kern 0.1em$9$}}
\def\Lyten{\hbox{{\rm Ly}\kern 0.1em$10$}}
\def\kms{\hbox{km~s$^{-1}$}}
\def\cmsq{\hbox{cm$^{-2}$}}
\def\cc{\hbox{cm$^{-3}$}}

\usepackage{soul}

\newcommand{\angstrom}{\mbox{\normalfont\AA}}
\definecolor{blacky}{rgb}{0,0,0}
\newcommand{\change}[1]{{\color{blacky}{#1}}}
\newcommand{\dn}[1]{{\color{blacky}{#1}}}

\usepackage{tabularx,ragged2e}
\usepackage{booktabs,array,times}

\title[Metal-Rich, Cool-Warm Gas in Galaxy Clusters]{Detection of Metal-Rich, Cool-Warm Gas in the Outskirts of Galaxy Clusters}

\author[Pradeep et al.]{Jayadev Pradeep,$^{1}$\thanks{E-mail: jayadev\_pradeep@yahoo.com} Anand Narayanan,$^{1}$ Sowgat Muzahid,$^{2}$ Daisuke Nagai,$^{3,4}$
\newauthor Jane C. Charlton,$^{5}$ and Raghunathan Srianand$^{6}$\\
$^{1}$Department of Earth and Space Sciences, Indian Institute of Space Science \& Technology, Thiruvananthapuram 695547, Kerala, INDIA\\
$^{2}$Leiden Observatory, Leiden University, P.O. Box 9513, NL-2300 RA Leiden, The Netherlands\\
$^{3}$Department of Physics, Yale University, New Haven, CT 06520, USA\\
$^{4}$Department of Astronomy, Yale University, New Haven, CT 06520, USA\\
$^{5}$Department of Astronomy \& Astrophysics, The Pennsylvania State University, University Park, PA 16802, USA\\
$^{6}$Inter-University Centre for Astronomy and Astrophysics, Post Bag 4, Ganeshkhind, Pune 411 007, INDIA
}

\date{Accepted 2019 July 23. Received 2019 July 8; in original form 2019 May 1}

\pubyear{2019}

\begin{document}
\label{firstpage}
\pagerange{\pageref{firstpage}--\pageref{lastpage}}
\maketitle

\begin{abstract}
We present an ultraviolet quasar absorption line analysis of metal lines associated with three strong intervening {\HI} absorbers (with $N(\HI)$ $>$ 10$^{16.5}$ cm$^{-2}$) detected in the outskirts of Sunyaev-Zel'dovich (SZ) effect-selected galaxy clusters ($z_{\rm cl} \sim 0.4 - 0.5$), within clustocentric impact parameters of $\rho_{cl}$ $\sim$ $(1.6 - 4.7)r_{500}$. \change{Discovered in a recent set of targeted far-UV $HST$/COS spectroscopic observations}, these absorbers have the highest {\HI} column densities ever observed in the outskirts of galaxy clusters, and are also rich in metal absorption lines. Photoionization models yield single phase solutions for the three absorbers with gas densities of $n_{\H} \sim 10^{-3} - 10^{-4}$~{\cc} and metallicities of [X/H] $>$ -1.0 (from one-tenth solar to near-solar). The widths of detected absorption lines suggest gas temperatures of $T \sim 10^4$~K. The inferred densities (temperatures) are significantly higher (lower) compared to the X-ray emitting intracluster medium in cluster cores. \change{The absorbers are tracing a cool phase of the intracluster gas in the cluster outskirts, either associated with gas stripped from cluster galaxies via outflows, tidal streams or ram-pressure forces, or denser regions within the intracluster medium that were uniformly chemically enriched from an earlier epoch of enhanced supernova and AGN feedback.}
\end{abstract}


%%%%%%%%%%%%%%%%%%%%%%%%%%%%%%%%%%%%%%%%%%%%%%%%%%

%%%%%%%%%%%%%%%%% BODY OF PAPER %%%%%%%%%%%%%%%%%%

\section{Introduction}\label{section:intro}

The intracluster medium (ICM) is the most dominant baryonic component of galaxy clusters, with the bulk ($70 - 80$\%) of the ICM consisting of hot ($T \gtrsim 10^6$~K) X-ray emitting plasma and the rest in cool-warm ($T < 10^6$~K) gas and stars (e.g., Ettori \hyperlink{Ettori2003}{2003}, Fukugita \& Peebles \hyperlink{Fukugita2004}{2004}, Kravtsov {\etal}\hyperlink{Kravtsov2005}{2005}, Gonzalez {\etal}\hyperlink{Gonzalez2007}{2007}, Planelles {\etal}\hyperlink{Planelles2013}{2013}). Until recently, the census of baryons in galaxy clusters has primarily been based on X-ray observations of the shock-heated ICM (e.g., White \& Rees \hyperlink{White1978}{1978}, Cen \& Ostriker \hyperlink{Cen1999}{1999}, Nagai \& Kravtsov \hyperlink{Nagai2003}{2003}, Ryu {\etal}\hyperlink{Ryu2003}{2003}). As a result, such studies have mostly been limited to well within the virialized regions ($\dn{r}\lesssim \change{r_{500}}$\footnote{$r_{500}$ is the over-density radius, defined as the cluster radius within which the enclosed mean total mass density is 500 times the critical density of the universe at the cluster redshift.}) of galaxy clusters, as the X-ray surface brightness of the ICM decreases radially from the cluster centre to the outskirts  (e.g., De Grandi \& Molendi \hyperlink{DeGrandi2002}{2002}, Vikhlinin {\etal}\hyperlink{Vikhlinin2006}{2006}).

Recent advances in X-ray and microwave observations have significantly extended measurements of the hot X-ray emitting gas into the outskirts of galaxy clusters (e.g., Simionescu {\etal}\hyperlink{Simionescu2011}{2011}, Walker {\etal}\hyperlink{Walker2013}{2013}, Urban {\etal}\hyperlink{Urban2017}{2017}, Mroczkowski {\etal}\hyperlink{Mroczkowski2019}{2019}, Walker {\etal}\hyperlink{Walker2019}{2019}). Modern cosmological hydrodynamic simulations show the outskirts of galaxy clusters as a dynamically active place. The cool-warm circumgalactic medium (CGM) of galaxies infalling into clusters are likely to get displaced from their galactic potential through ram-pressure forces exerted by the hot ICM. The gas thus \change{removed} can be present at distances much beyond the cluster virial radius (DeGrandi {\etal}\hyperlink{DeGrandi2016}{2016}). 
\change{The Virgo cluster offers a local example of cool gas stripped from two galaxies (M86 and NGC 4438) during a sub-cluster tidal interaction between them, resulting in a spectacular complex of H$\alpha$ filaments permeating the ICM (Kenney {\etal}\hyperlink{Kenney2008}{2008}, Ehlert {\etal}\hyperlink{Ehlert2013}{2013}). A similar complex of H$\alpha$-emitting intracluster filaments has been observed at the center of the Perseus cluster, inferred to be due to the interaction of NGC 1275 with a group of gas-rich galaxies (Fabian {\etal}\hyperlink{Fabian1984}{1984}, Conselice {\etal}\hyperlink{Conselice2001}{2001}). The gas displaced away from galaxies
%by ways of different
\dn{through various} mechanisms gets mixed with the surrounding ICM, creating an inhomogeneous (\change{Nagai \& Lau \hyperlink{Nagai2011}{2011}}, Vazza {\etal}\hyperlink{Vazza2013}{2013}, Zhuravleva {\etal}\hyperlink{Zhuravleva2013}{2013}, Rasia {\etal}\hyperlink{Rasia2014}{2014}) and turbulent (Lau {\etal}\hyperlink{Lau2009}{2009}, Nelson {\etal}\hyperlink{Nelson2014}{2014}) medium in the outskirts of clusters}. Such non-linear astrophysical processes, if not understood and modelled properly, can lead to significant systematic uncertainties in the cosmological constraints derived from X-ray and microwave observations of galaxy clusters (see Pratt {\etal}\hyperlink{Pratt2019}{2019} for a recent review). Finding observational signatures of such gas in cluster environments is therefore crucial.

Relatively metal-poor cool-warm gas can also penetrate into galaxy clusters through gas streams from the cosmic web of intergalactic filaments (Zinger {\etal}\hyperlink{Zinger2016}{2016}). Cosmological simulations also show the mass fraction of the cool-warm gas as increasing with the cluster-centric radius, becoming comparable to or greater than the hot gas mass fraction at $r\gtrsim \change{3r_{500}}$ (Emerick {\etal}\hyperlink{Emerick2015}{2015}, Butsky {\etal}\hyperlink{Butsky2019}{2019}), implying that high neutral column density gas, yet to be subject to cluster virial shocks, must be traceable in the outskirts of galaxy clusters. Cumulatively, these rich and complex dynamical processes eventually gives rise to a multiphase ICM with a range of physical and chemical properties in the cluster outskirts (Butsky {\etal}\hyperlink{Butsky2019}{2019}).

Observationally, unlike the hot X-ray emitting gas, the thermodynamic, kinematic and chemical properties of the cool-warm gas in cluster outskirts remain less explored. Quasar absorption line spectroscopy serves as a suitable probe of such multiphase gas, especially in the outskirts of galaxy clusters where they cannot be seen in emission. There have been only a handful of absorption line spectroscopic studies targeted at the ICM and CGM in the outskirts of galaxy clusters.
Yoon {\etal}(\hyperlink{Yoon2012}{2012}) identified several {\Lya} absorbers with {\HI} column densities of $N(\HI) \lesssim 10^{15.5}$ cm$^{-2}$
%$\log~N(\HI) \lesssim 15.5$ 
probing gas with $T = 10^4 - 10^5$~K in the Virgo cluster environment. The absorbers were in regions distinct from the hot ICM, with the covering fraction of {\HI} showing an increase at distances beyond the virial radius. Comparable results were also obtained by Yoon {\etal}(\hyperlink{Yoon2017}{2017}) for absorbers associated with the Coma cluster. Similar examples of {\HI} absorbers tracing cooler intracluster gas were also presented by Burchett {\etal}(\hyperlink{Burchett2018}{2018}), which they interpreted as gas infall from the cosmic web. On the other hand, Manuwal {\etal}(\hyperlink{Manuwal2019}{2019}) interpreted the presence of cool supersolar metallicity gas in the outskirts of Virgo cluster as possibly interstellar gas displaced from galaxies through outflows or tidal interactions. 

%In this work we use the absorption line-spectroscopic technique in the ultraviolet (UV) to study the cool-warm gas phase in three $z \sim 0.5$ galaxy clusters, complementing the more conventional X-ray and microwave observations of cluster gas mentioned earlier. 

Motivated by the small number of targeted studies of the ICM and CGM in galaxy cluster outskirts, Muzahid {\etal}(\hyperlink{Muzahid2017}{2017}; hereafter M17), carried out a pilot program using $HST$/COS of lines of sight towards background UV-bright quasars that probe cluster outskirts. The far-UV spectroscopic data towards three different \change{SZ-selected} clusters at $z \sim 0.46$ revealed the presence of large columns of {\HI} gas ($N(\HI) > 10^{16.5}$ cm$^{-2}$) \change{at clustocentric impact parameters beyond $1.5r_{500}$}. \change{The three {\HI} absorbers are at redshifts $z =$ $0.43737$, $0.43968$ \& $~0.51484$, with the associated galaxy clusters having photometric redshifts of $z_{cl} =$ $0.45$, $0.45$ \& $0.47$, respectively (Bleem {\etal}\hyperlink{Bleem2015}{2015}). The clusters have  estimated masses of 3.04 $\times$ 10$^{14}$ M$_{\odot}$, 3.19 $\times$ 10$^{14}$ M$_{\odot}$ and 3.81 $\times$ 10$^{14}$ M$_{\odot}$ and $r_{500}$ values of 0.87 Mpc, 0.89 Mpc and 0.93 Mpc (Bleem {\etal}\hyperlink{Bleem2015}{2015}). The absorbers are at projected separations of 3.8~Mpc ($\change{\rho_{cl}}/r_{500} = 4.4$), 4.2~Mpc ($\change{\rho_{cl}}/r_{500} = 4.7$) and 1.5~Mpc ($\change{\rho_{cl}}/r_{500} = 1.6$) from the respective cluster centers, away from the hot and tenuous central X-ray emitting regions. The properties of these QSO-cluster pairs are \dn{listed} in Table \ref{table:qsocluster}.} The {\HI} column densities for these absorbers are one of the highest ever measured for the diffuse gas in galaxy clusters (also see Tripp {\etal}\hyperlink{Tripp2005}{2005}), causing full or partial Lyman limit breaks in background quasar spectra. 
%The absorbers are unlikely to be tracing circumgalactic material, as the covering fraction of such gas is expected to be low ($\lesssim 18$\%, for {\HI} gas with $W_r(\mathrm{Ly}\alpha) > 30$~m{\AA}) around galaxies that are members of clusters compared to field galaxies (\hyperlink{Burchett2018}{Burchett {\etal}2018}). 
Based on the analysis of the Lyman series lines in each absorber, M17 concluded that they are tracing $T \sim 10^4$~K gas. The effective $b$-parameters of the {\HI} lines were less than the typical subsonic random gas motions ($\sigma_{gas}\approx 300$ km/s) expected in the hot X-ray emitting ICM in galaxy clusters suggested by simulations (e.g., Nagai {\etal}\hyperlink{Nagai2013}{2013}). The origin of such large amounts of cool gas observed in these massive galaxy clusters thus remain uncertain. 

\begin{table*}\centering
\setlength{\tabcolsep}{15pt}
\renewcommand{\arraystretch}{1.5}
	\begin{tabular}{cccccccc}
		\hline
		Cluster    & z$_{cl}$       & M$_{500}$      & r$_{500}$ & QSO        & z$_{QSO}$  & $\rho_{cl}/r_{500}$  & z$_{abs}$                            \\
		&           & ($10^{14}$ M$_{\odot}$) & Mpc  &      & Mpc   &            &                                  \\
		(1)        & (2)       & (3)       & (4)  & (5)        & (6)   & (7)    & (8)                               \\ \hline
		J0041-5107 & 0.45$ \pm$ 0.04 & 3.04 $\pm$ 0.87 & 0.87 & J0040-5057 & 0.608 & 4.4  &  0.43737                     \\
		J2016-4517 & 0.45 $\pm$ 0.03 & 3.19 $\pm$ 0.89 & 0.89 & J2017-4516 & 0.692 & 4.7  &  0.43968                     \\
		J2109-5040 & 0.47 $\pm$ 0.04 & 3.81 $\pm$ 0.87 & 0.93 & J2109-5042 & 1.262 & 1.6  &  0.51484  \\ \hline
	\end{tabular}
	\caption{Information about the QSO-cluster pairs. Cluster names (1), photometric redshifts (2), and masses (3) are from Bleem {\etal}(\protect\hyperlink{Bleem2015}{2015}). QSO names (5) and redshifts (6) are from Monroe {\etal}(\protect\hyperlink{Monroe2016}{2016}). The $r_{500}$ values (4), normalized clustocentric impact parameters of the QSO sightlines (7) and absorber redshifts (8) are from Muzahid {\etal}(\protect\hyperlink{Muzahid2018}{2018}).}
	%\caption{Information about the QSO-cluster pairs. Cluster names (1), photometric redshifts (2), and masses (3) are from Bleem {\etal}(\hyperlink{Bleem2015}{2015}). QSO names (5) and redshifts (6) are from Monroe {\etal}(\hyperlink{Monroe2016}{2016}). The $r_{500}$ values (4), normalized clustocentric impact parameters of the QSO sightlines (7) and absorber redshifts (8) are from Muzahid {\etal}(\hyperlink{Muzahid2018}{2018}).}
	\label{table:qsocluster}
\end{table*}

In this work, we have analyzed the metal lines associated with the three high column density cluster absorbers reported in M17. None of the previous studies have done a comprehensive metal line analysis of absorbers associated with the cooler phase of the ICM in the cluster outskirts. The different line widths of low ionization metal lines combined with {\HI} provide an estimate on the temperature of the gas phase without making explicit assumptions about the line broadening mechanism. Through ionization modelling of these absorbers, we determine the metallicity and relative chemical abundances, which are crucial for interpreting the nature and origin of these clouds.

This article is divided into six sections. In Section~\ref{section:data}, we briefly describe the HST/COS data that have been used. Section~\ref{section:measurements} describes each of the absorption systems and the analysis of the associated metal lines. The photoionization modelling results and derived physical conditions in the absorbers are presented in Section~\ref{section:modelling}. We discuss the possible astrophysical origin of these absorbers in Section~\ref{section:discussion} and summarize our conclusions in Section~\ref{section:conclusions}. Throughout the work we have adopted a flat $\Lambda$CDM cosmology with H$_0$ = 71 km s$^{-1}$ Mpc$^{-1}$, $\Omega_{M}$ = 0.3, and $\Omega_{\Lambda}$ = 0.7. 

\section{DATA}\label{section:data}

The far-UV COS spectra for the three QSO sightlines UVQS J0040-5057, UVQS J2017-4516 and UVQS J2109-5042 were obtained as part of the GO program ID: 14655 (PI: Muzahid). The reduced spectra were obtained from the \textit{Hubble Spectroscopic Legacy Archive} (Peeples {\etal}\hyperlink{Peeples2017}{2017}). The G130M and G160M grating data offer a combined wavelength coverage from 1100 - 1800 $\angstrom$ with a signal-to-noise ratio of $S/N \sim 10$ per resolution element after Nyquist sampling. Low order polynomials were used to fit the local continuum, avoiding evident absorption features. Line measurements were carried out on the continuum-normalized spectra. The integrated column densities were measured using the apparent optical depth (AOD) method of Savage \& Sembach (\hyperlink{Savage1991}{1991}) which offers a convenient means for converting velocity-resolved flux profiles of unsaturated lines into column density measurements. For saturated lines, the AOD method provides a lower limit on the column density. Voigt profile fitting was also performed on these lines using the VPFIT routine (ver.10, Kim {\etal}\hyperlink{Kim2007}{2007}) after convolving the observed profiles with the COS theoretical FUV instrumental line spread functions given by Kriss (\hyperlink{Kriss2011}{2011}) and on the STScI website\footnote{http://www.stsci.edu/hst/cos/performance/spectral\_resolution/}. 
%Only those lines with a significance of $\geq 3\sigma$ are considered as detected. 
\change{For non-detections, useful column density upper limits are obtained from the 3$\sigma$ equivalent width uncertainties, using the linear regime of the Curve of Growth}. We adopt the same redshifts for the absorbers as given in M17. 

%obtained by integrating over the same velocity range as the corresponding {\CII} 1036 lines for the z$_{abs}$ $=$ 0.43737 and z$_{abs}$ $=$ 0.43968 absorbers, and the {\CIII} 977 line in the case of the z$_{abs}$ $=$ 0.51484 absorber, and determining the corresponding column density in the linear part of the curve-of-growth (COG). 

\section{System Description and Measurements}\label{section:measurements}

\subsection{The z$_{abs}$ $=$ 0.43737 absorber towards UVQS J0040-5057}

\begin{figure*}
	\centering
	\includegraphics[scale=0.20]{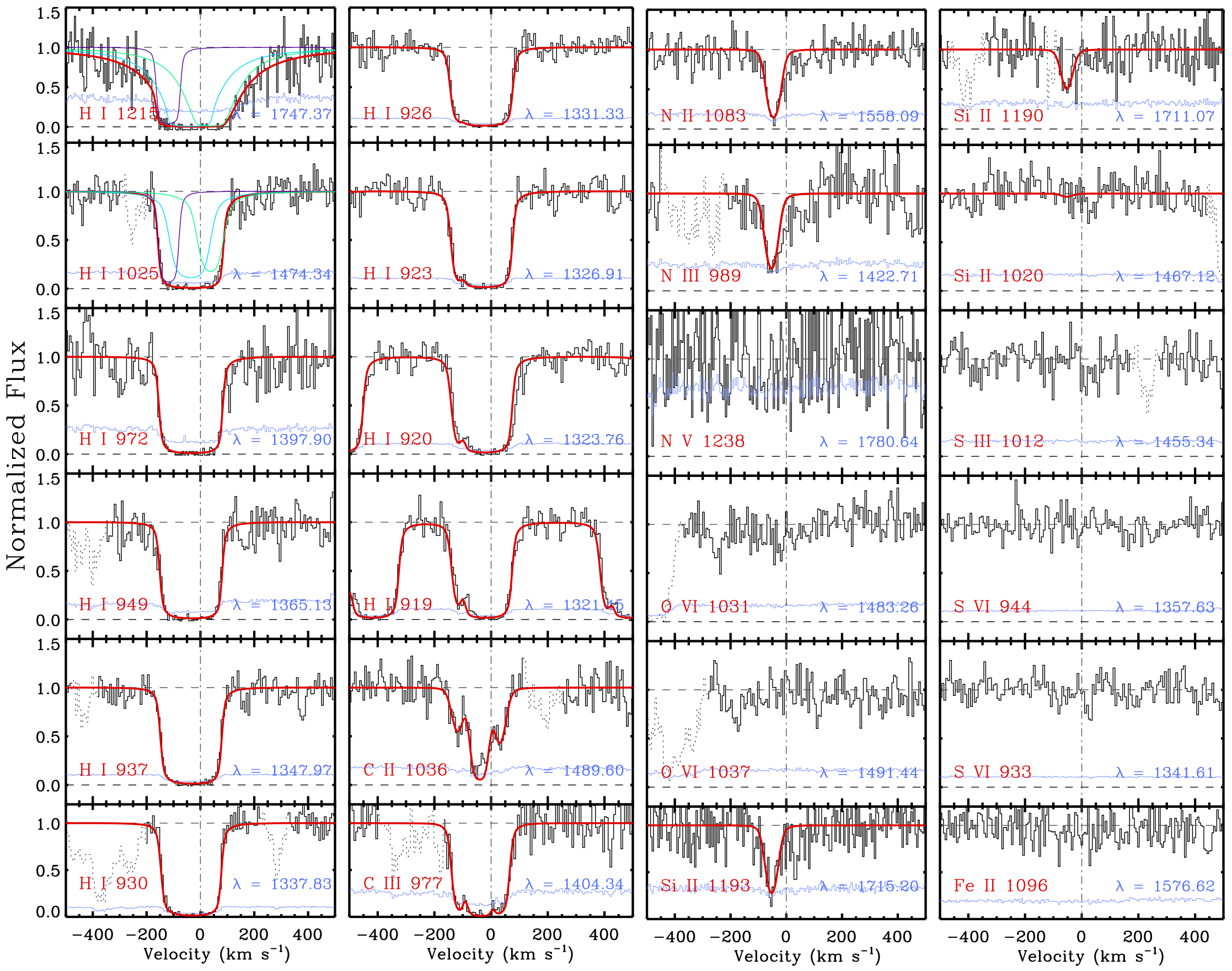}
	\caption{Velocity plots of the z$_{abs}$ $=$ 0.43737 absorber toward UVQS J0040-5057, with continuum-normalized flux along the Y-axis and the velocity scale relative to the redshift of the absorber along the X-axis. $v = 0$~{\kms}, marked by the \textit{dashed-dotted} vertical line, indicates the absorber redshift. The $1\sigma$ uncertainty in flux is indicated by the \textit{blue} curve at the bottom of each panel. The \textit{red} curves are the best-fit Voigt profiles. For Ly-$\alpha$ and Ly-$\beta$, the contributions from the separate components are also shown.\label{fig:sys1plot}}
\end{figure*}

The system plot for the absorber is shown in Figure \ref{fig:sys1plot}, and the AOD and Voigt profile fit measurements are listed in Table \ref{table1}. This is a Lyman limit absorber with very strong associated metal absorption lines. Using a single component COG analysis on the available Lyman series lines, M17 estimate the atomic hydrogen column density as $N(\HI)$ $=$ $10^{18.63~{\pm}~0.15}$~{\cmsq}, consistent with the weak damping wing seen in {\Lya}. The COG column density is also in agreement with the presence of a full Lyman limit break in the observed spectrum at $\lambda < 1310$~{\AA}. The {\CII}~$1036$ and {\CIII}~$977$ metal lines indicate a three-component structure to the absorption, which is also partially evident in the higher order Lyman lines. Taking a hint from this, we have simultaneously fitted a three-component model to the {\HI} lines. The best-fit model parameters (refer Table \ref{table1}) also yield a total {\HI} column density of $N(\HI) = 10^{18.83~{\pm}~0.20}$~{\cmsq}, agreeing with the COG measurement given in M17. The components at $v \sim -38$~{\kms} and $+36$~{\kms} are saturated even in the higher order Lyman lines, and therefore the fit results for these two components may not be unique. However, the column densities of these two components cannot be significantly larger than what we measure, as that would require the two {\HI} components to be narrower than the corresponding metal lines. 

 The {\CII} absorption in the central ($v \sim -38$~{\kms}) component is saturated. The column density estimation for this component is likely to be less certain than the $0.11$~dex uncertainty obtained from profile fitting. The {\CIII} line also suffers from a high degree of saturation at the core. Based on the profile fit results, we expect the true column density to be $N(\CIII) \gtrsim 10^{15.3}$~{\cmsq}, which is the value we adopt for constraining the ionization models.

%For each pixel, the velocity $v$ is calculated by the relation $v = c \times (\frac{(1 + z_{pixel})^2 - (1 + z_{abs})^2}{(1 + z_{pixel})^2 + (1 + z_{abs})^2}$, where $z_{pixel}$ of the pixel is calculated from the observed wavelength of the pixel ($\lambda_{pixel}$) and the rest wavelength ($\lambda_{rest}$) of the transition as $ z_{pixel} = \lambda_{pixel}/\lambda_{rest} - 1$.

%\begin{figure*}
%	\centering
%	\includegraphics[scale=0.30]{figs/sysplot_sys1_2.png}
%	\caption{System plot of the z$_{abs}$=0.43737 absorber toward UVQS J0040-5057, continued from Figure \ref{fig:sys1plot1}. \label{fig:sys1plot2}}
%\end{figure*}

The {\NII}, {\NIII} and {\SiII} lines do not show a component structure but the absorption seems to be arising from the $v \sim -38$~{\kms} component  corresponding to the strongest {\Lya}, {\CII} and {\CIII} absorption. Though the {\SiII}~$1193$ line is strong, simultaneously fitting it with the weaker {\SiII}~$1190$ and the non-detected {\SiII}~ $1020$ lines, yield a reliable measurement on the column density and the $b$-parameter. The {\NII}~$1083$ and {\NIII}~$989$ lines were fitted by allowing their Doppler parameters to vary together, assuming that the two species are arising from the same phase. However, these two lines are strong and possibly saturated and hence the corresponding measurement of $N(\NII)$ and $N(\NIII)$ are taken as lower limits.  The {\OVIdblt} doublet is covered, but not detected, indicating \change{\dn{low-ionization gas with prevalence} of photoionization}. Coverage of {\CIV} and {\SiIV} would be needed to probe the presence of multiple ionization phases. 

\subsection{The z$_{abs}$ $=$ 0.43968 absorber towards UVQS J2017-4516}

\begin{figure*}
	\centering
	\includegraphics[scale=0.20]{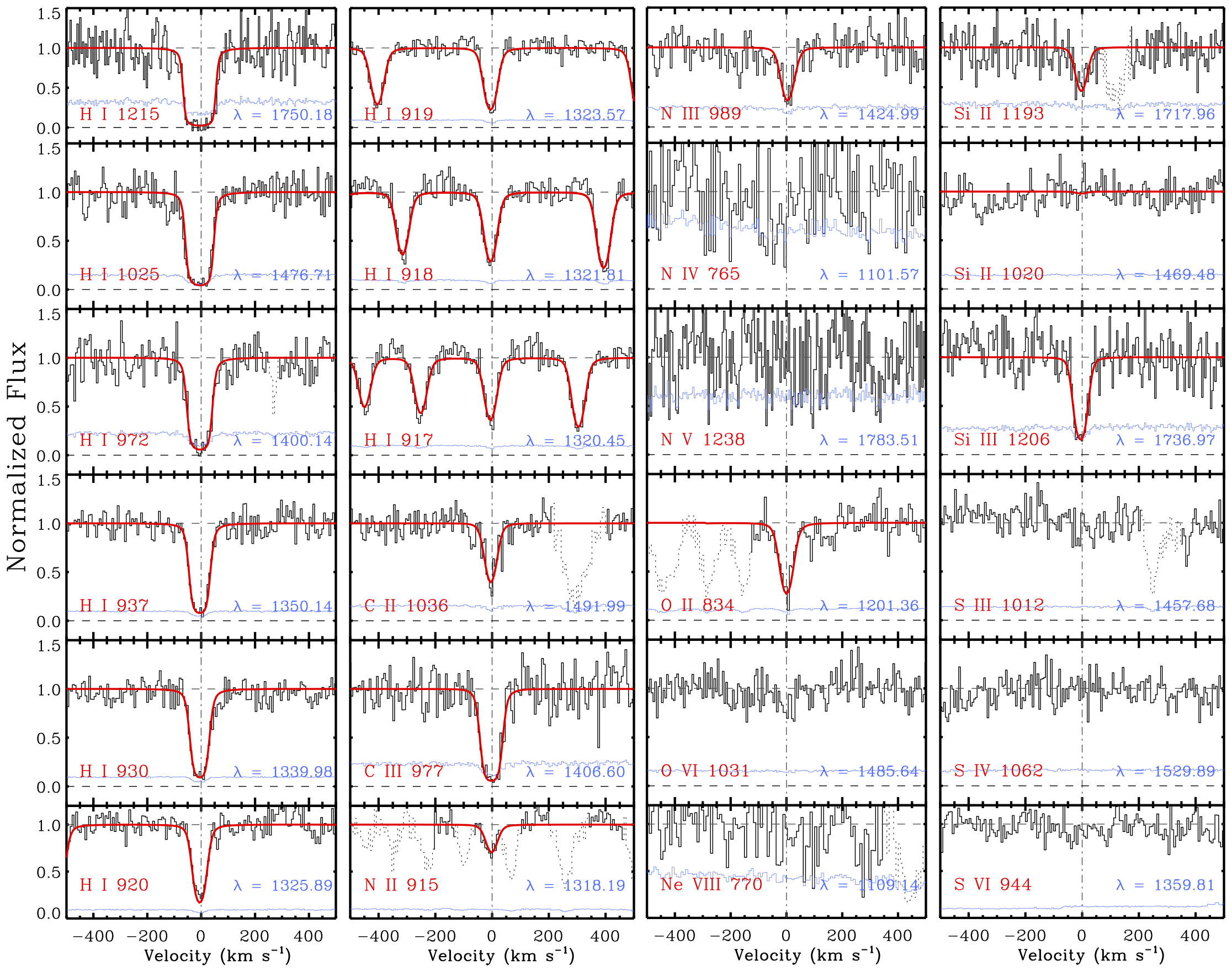}
	\caption{Velocity plots of the z$_{abs}$ $=$ 0.43968 absorber toward UVQS J2017-4516, with continuum-normalized flux along the Y-axis and the velocity scale relative to the redshift of the absorber along the X-axis. The $1\sigma$ uncertainty in flux is indicated by the \textit{blue} curve at the bottom of each panel. The \textit{red} curves are the best-fit Voigt profiles.\label{fig:sys2plot}}
\end{figure*}

This is a partial Lyman limit absorber with  $N(\HI)$ $=$ $10^{16.52~{\pm}~0.15}$~{\cmsq}, obtained by M17 from a COG analysis of the {\HI} Lyman transitions lines. The unsaturated higher order Lyman lines have simple symmetrical profiles that are well explained by a single-component model fit. The system plot for this absorber is shown in Figure \ref{fig:sys2plot}, and the line measurements are listed in Table \ref{table2}. The {\CII}~$1036$, {\NII}~$915$, {\NIII}~$989$, {\OII}~$834$ and {\SiII}~$1193$ lines are unsaturated and offer the best constrained column densities. The {\CIII}~$977$ and {\SiIII}~$1206$ lines have saturated cores. The $b$ parameters of metal lines are expected to be similar as they are from ions of similar mass tracing the same gas phase. We therefore fitted the metal lines by allowing the $b$ parameters to vary together, which helps the fitting routine to compensate for line saturation to an extent. 
%The {\OVI}~$1031$ appears as a weak feature with a formal detection significance of $3.5\sigma$ obtained by integrating the spectrum over the same velocity range as other metal lines. 
The non-detection of {\OVI}~$1037$ is consistent with the weak detection of the {\OVI}~$1031$ line. The spectrum also covers {\NeVIII}~$770$ and {\SVI}~$944$ which are non-detections. \change{While the meagre {\OVI} detection is indicative of a possible origin via collisional ionization, the non-detection of other higher ionization lines and the presence of strong low-ionization absorption imply \dn{predominance} of photoionized gas}. The metal line widths are comparable with the {\HI} line width, indicating significant non-thermal contribution to the line broadening, with the neutral hydrogen \textit{b}-parameter suggesting an upper limit of $T = 3.2 \times 10^4$~K for the gas temperature.

%\begin{figure*}
%	\centering
%	\includegraphics[scale=0.20]{figs/system2.png}
%	\caption{Velocity plots of the z$_{abs}$=0.43968 absorber toward UVQS J2017-4516, with continuum-normalized flux along the Y-axis and the velocity scale relative to the redshift of the absorber along the X-axis. The $1\sigma$ uncertainty in flux is indicated by the \textit{blue} curve at the bottom of each panel. The \textit{red} curves are the best-fit Voigt profiles.\label{fig:sys2plot}}
%\end{figure*}

%\begin{figure*}[t]
%	\centering
%	\includegraphics[scale=0.30]{figs/sysplot_sys2_2.png}
%	\caption{System plot of the z$_{abs}$=0.43968 absorber toward UVQS J2017-4516, continued from Figure \ref{fig:sys2plot1}. \label{fig:sys2plot2}}
%\end{figure*}

\subsection{The z$_{abs}$ $=$ 0.51484 absorber towards UVQS J2109-5042}

\begin{figure*}
	\centering
	\includegraphics[scale=0.20]{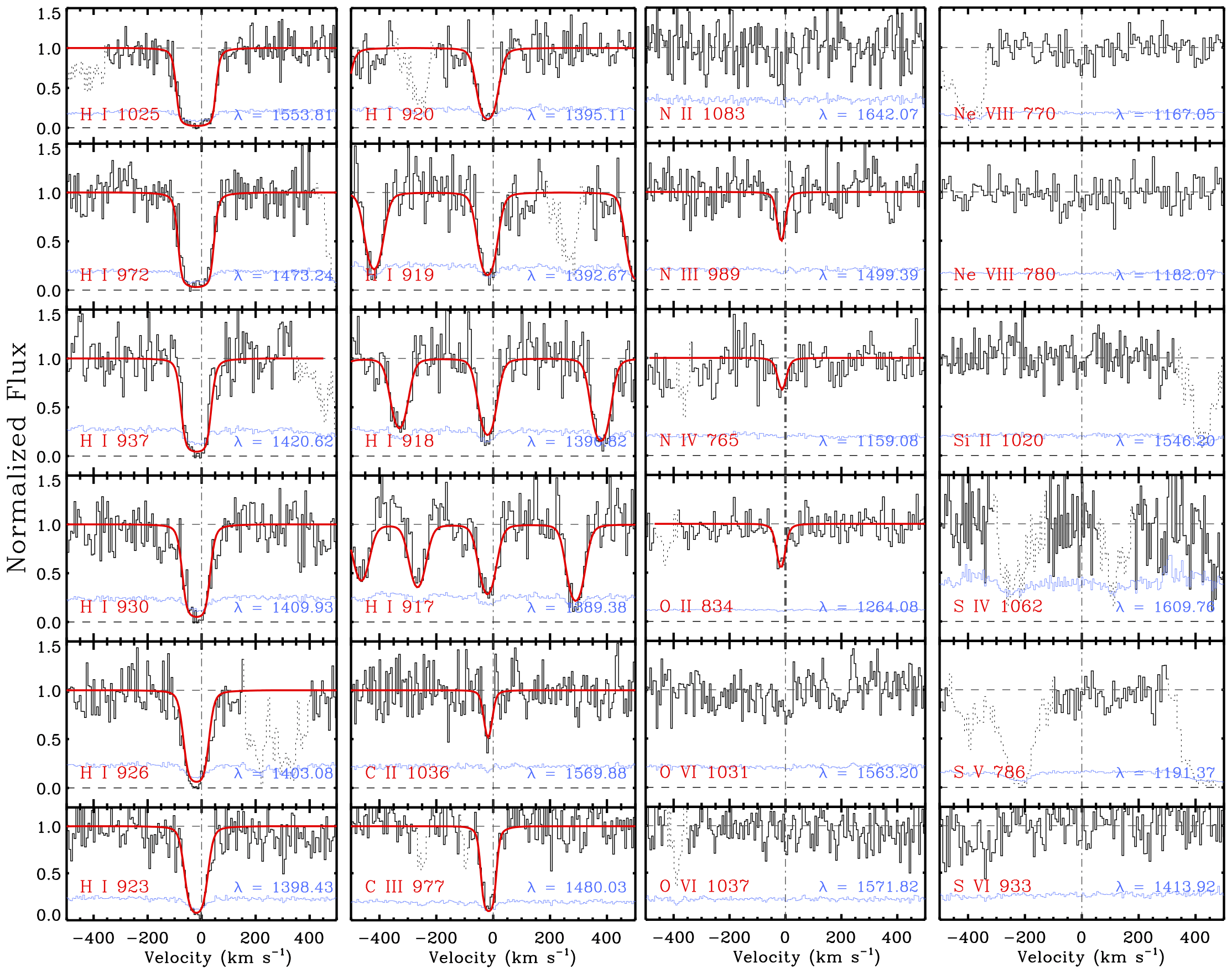}
	\caption{Velocity plots of the z$_{abs}$ $=$ 0.51484 absorber towards UVQS J2109-5042, with continuum-normalized flux along the Y-axis and the velocity scale relative to the redshift of the absorber along the X-axis. The $1\sigma$ uncertainty in flux is indicated by the \textit{blue} curve at the bottom of each panel. The \textit{red} curves are the best-fit Voigt profiles.\label{fig:sys3plot}}
\end{figure*}

With an $N(\HI)$ $=$ $10^{16.68~{\pm}~0.03}$~{\cmsq}, as obtained by M17 from a COG analysis to the neutral hydrogen Lyman transitions lines, this absorber also shows a partial Lyman break. The unsaturated higher order Lyman lines are well explained by a single-component model fit. The system plot for this absorber is shown in Figure \ref{fig:sys3plot}, and the line measurements are tabulated in Table \ref{table3}.

The detected {\CII}, {\CIII}, {\NIII}, {\NIV} and {\OII} absorption lines are fitted using single component Voigt profiles, while tying their Doppler parameters to vary in tandem under the assumption that these species arise from the same gas phase. 
The {\CIII} line is suggestive of core saturation, and therefore the measured $N(\CIII)$ should be taken as a lower limit, whereas the other detected metal lines, viz. {\CII}~$1036$, {\NIII}~$989$, {\NIV}~$765$ and {\OII}~$834$ are unsaturated, and the {\SiII}~$1020$ is a non-detection. The different $b$-parameters for {\HI} and metal lines solve for a temperature of $T = 4.7 \times 10^4$~K with $\sim 40$\% of the {\HI} line broadening coming from turbulence. The spectrum covers {\OVI}~$1031$, {\NeVIII}~$770$ and {\SVI}~$933$, which are all non-detections, implying a single phase with low ionization conditions, \change{dominated by photoionization}.   

%\begin{figure*}
%	\centering
%	\includegraphics[scale=0.30]{figs/sysplot_sys3_2.png}
%	\caption{System plot of the z$_{abs}$ $=$ 0.51484 absorber towards UVQS J2109-5042, continued from Figure \ref{fig:sys3plot1}. \label{fig:sys3plot2}}
%\end{figure*}
% Ne VIII 770 is 2 sigma detection

% B parameters of C II 1036, N III 989 and C III 977 are tied to O II  834 

\section{Physical Conditions and Chemical Abundances of the Absorbers} \label{section:modelling}

Photoionization modelling using Cloudy (Ferland {\etal}\hyperlink{Ferland2013}{2013}) was used to derive the physical condition and chemical abundances in the absorbers. These models assume that the gas clouds to be isothermal with constant density, plane parallel geometry and no dust content. The ionization in the cloud is assumed to be dominated by photoionization by the extragalactic UV background radiation at the absorber redshifts, for which we have used the model given by Khaire \& Srianand (\hyperlink{Khaire2019}{2019}; fiducial Q18 model, hereafter KS18). Assuming the solar relative elemental abundances of Asplund {\etal}(\hyperlink{Asplund2009}{2009}), photoionization simulations were run for the observed values of {\HI} column density, and gas densities ranging from $10^{-6}$ to $10^{-1}$ cm$^{-3}$. A suite of ionization models were generated by varying metallicities from [X/H] = $-2$ to [X/H] = $0.5$, to arrive at phase solutions that best explain the observed line measurements for the three absorbers. 

\subsection{The z$_{abs}$ $=$ 0.43737 absorber towards UVQS J0040-5057.}

\begin{figure}
     \centering
     \begin{subfigure}
         \centering
         \includegraphics[trim=2cm 2cm 1cm 2cm,scale=0.35]{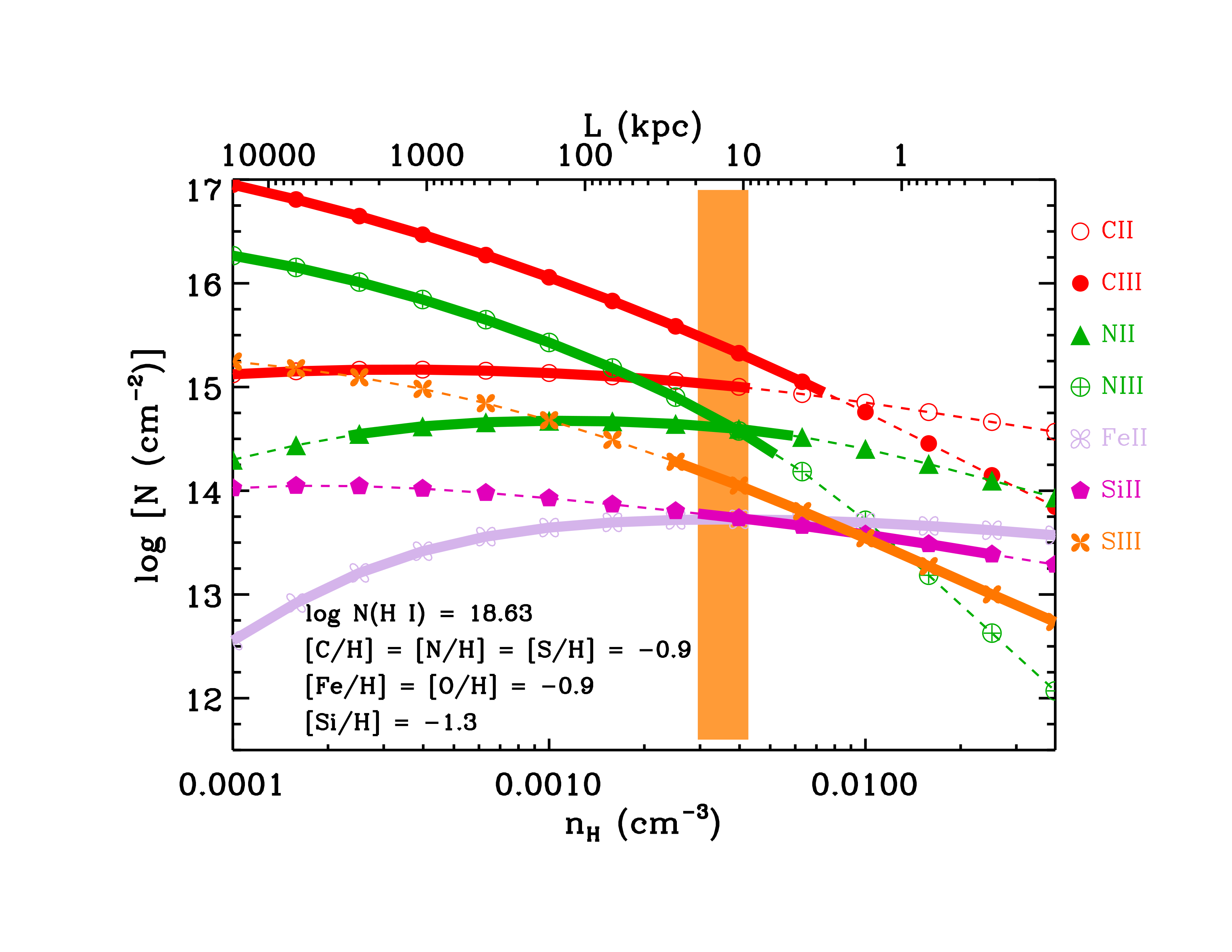}
     \end{subfigure}
     \begin{subfigure}
         \centering
         \includegraphics[trim=2cm 2cm 1cm 2cm,scale=0.35]{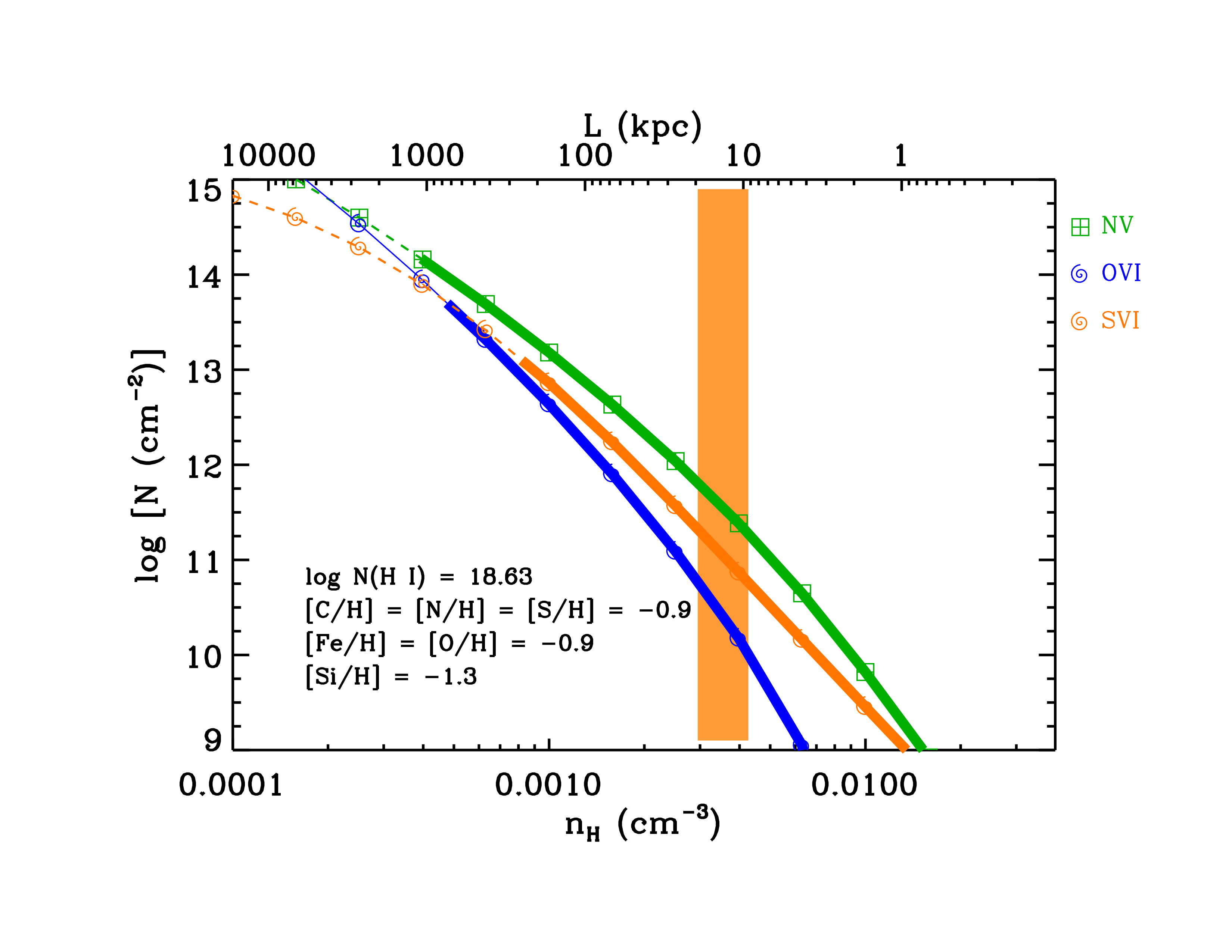}
     \end{subfigure}
     \caption{A possible photoionization equilibrium model for the z$_{abs}$=0.43737 absorber toward UVQS J0040-5057. The model assumes an abundance of -1.3 solar for silicon and -0.9 solar for other elements. The model-predicted variation of the column densities of various species (thin lines), along with the observed values (thick lines), are plotted against gas density (n$_{\H}$). Ions of low and intermediate ionizations occupy the \textit{top} panel, and high ions the \textit{bottom} panel. The narrow range of densities for which the observed column densities of ions are feasible, under the chosen chemical abundances, is marked with the \textit{orange} strip. The top X-axis shows the absorber line of sight thickness.}
     \label{fig:sys1}
\end{figure}

The photoionization models for this absorber are shown in Figure \ref{fig:sys1}. Though the kinematics of the {\CII} and {\CIII} lines suggest a three-component structure, our models are derived for integrated column densities. To account for saturation, we have considered {\CII}, {\CIII}, {\NII} and {\NIII} column densities as lower limits, whereas the unsaturated {\SiII} is treated as a measurement. 

From varying the metallicity, it is found that the observed $N(\SiII)$ and the lower limits on $N(\CII)$ and $N(\NII)$ cannot be explained for any gas density for abundances of [Si/H] $\leq -1.9$, [C/H] and [N/H] $< -0.9$. These lower limits on abundances are true for $n_{\H} = 5 \times 10^{-4}$~{\cc}, where the ionization fraction of {\SiII}, {\CII} and {\NII} peak. A single phase solution for these limiting abundances and density is also consistent with the observed {\CIII} lower limit and the non-detections of other metal lines. However, it requires a deviation of [C/Si] relative abundance from solar of 0.3 dex. Single phase solutions are also feasible for higher abundances of silicon ([C/Si]~$\lesssim 0.3$) at densities of $n_{\H} > 5 \times 10^{-4}$~{\cc}. One such solution is illustrated in Figure \ref{fig:sys1}. For $n_{\H} > 5 \times 10^{-4}$~{\cc}, the models also predict a total hydrogen column density of $N(\H) \lesssim 7.6 \times 10^{20}$~{\cmsq}, a thermal gas pressure of $p/K \gtrsim 8.62$~K~{\cc}, and a line of sight thickness of $L \lesssim 492.8~kpc$. The temperatures of $T \lesssim 1.7 \times 10^{4}$, predicted by the models, is supported by the metal line widths with non-thermal factors contributing $\gtrsim 70\%$ of line broadening.

\subsection{The z$_{abs}$ $=$ 0.43968 absorber towards UVQS J2017-4516}

\begin{figure*}
     \centering
     \begin{subfigure}
         \centering
         \includegraphics[trim=2cm 2cm 1cm 2cm,scale=0.35]{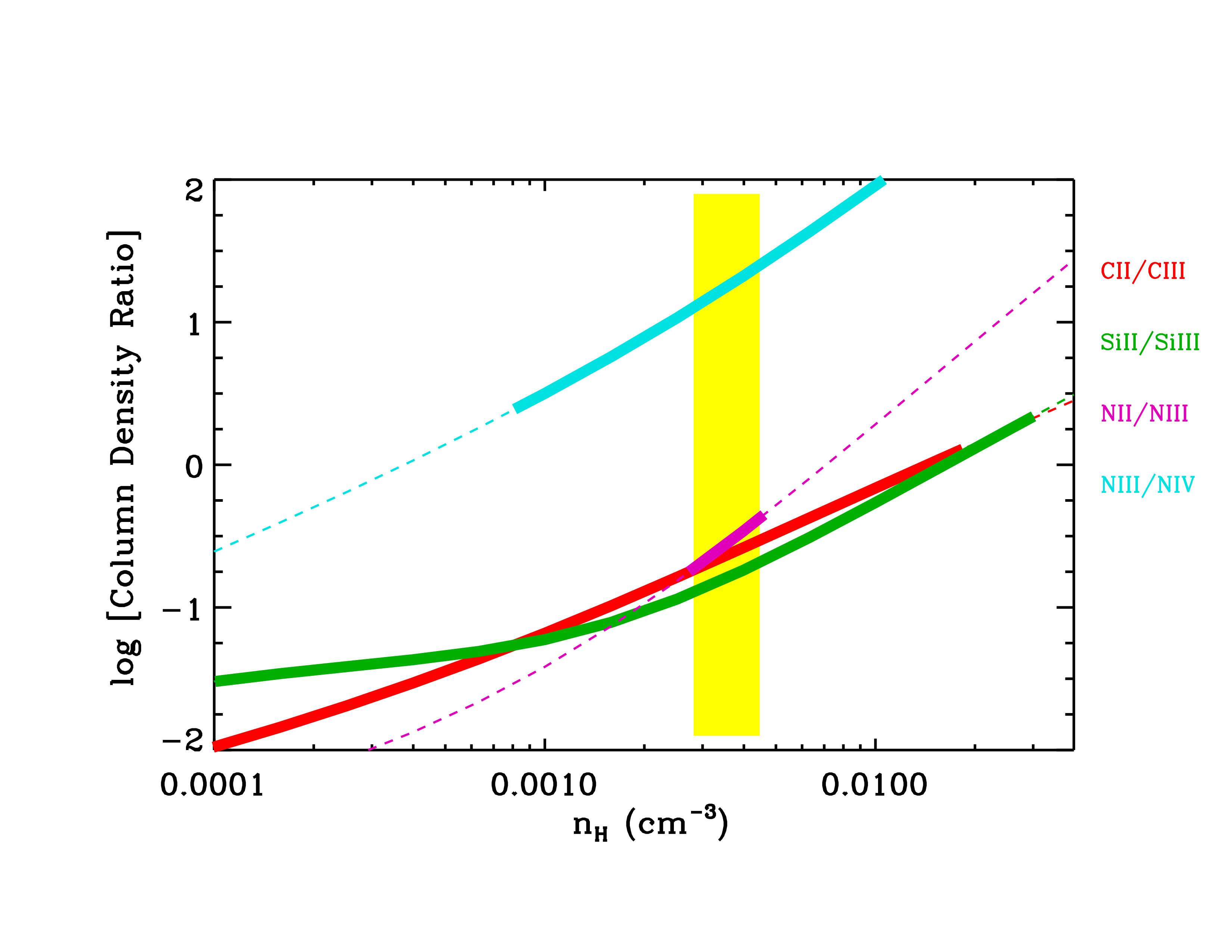}
     \end{subfigure}%
     \begin{subfigure}
         \centering
         \includegraphics[trim=2cm 2cm 2cm 2cm,scale=0.35]{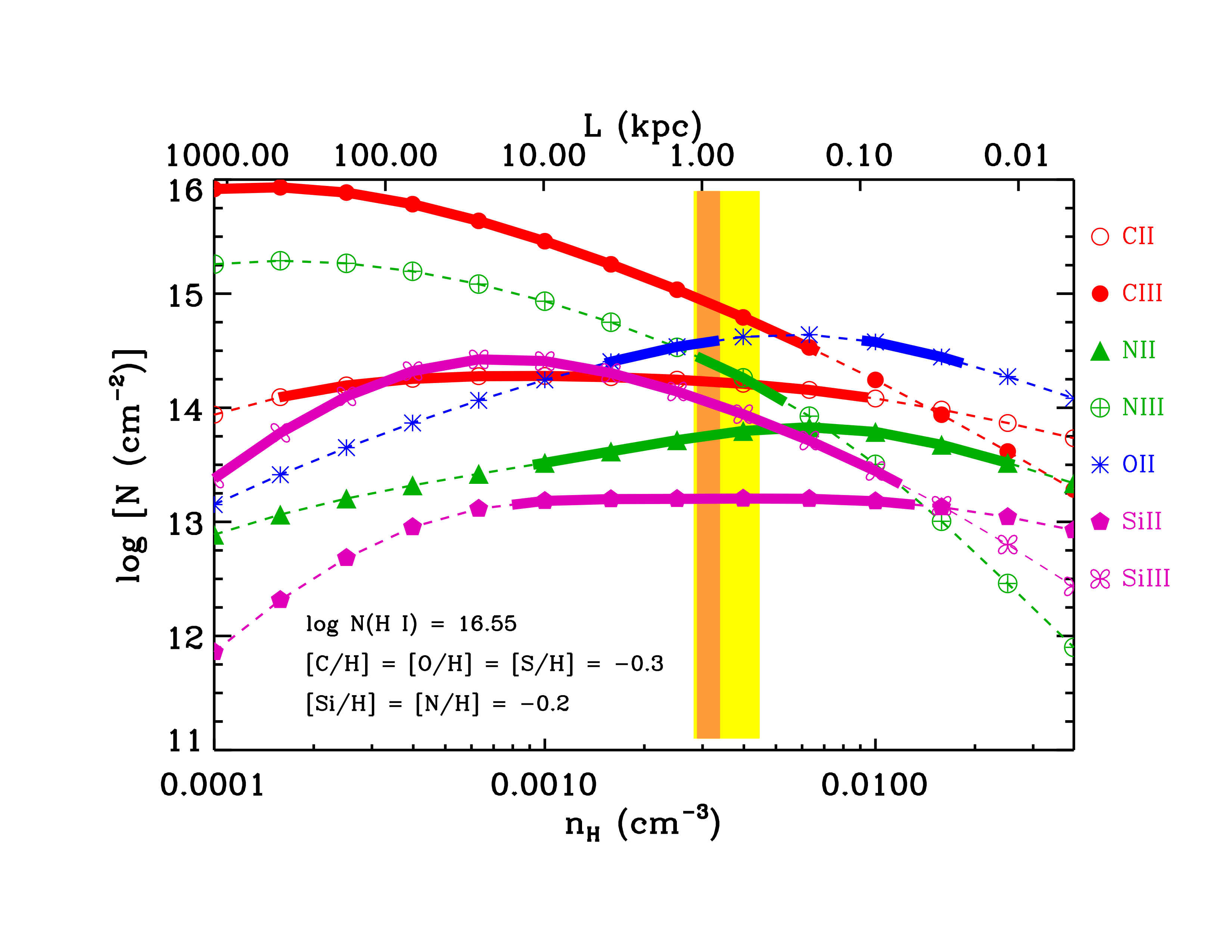}
     \end{subfigure}
     \begin{subfigure}
         \centering
         \includegraphics[trim=3.5cm 2cm 0.5cm 1cm,scale=0.35]{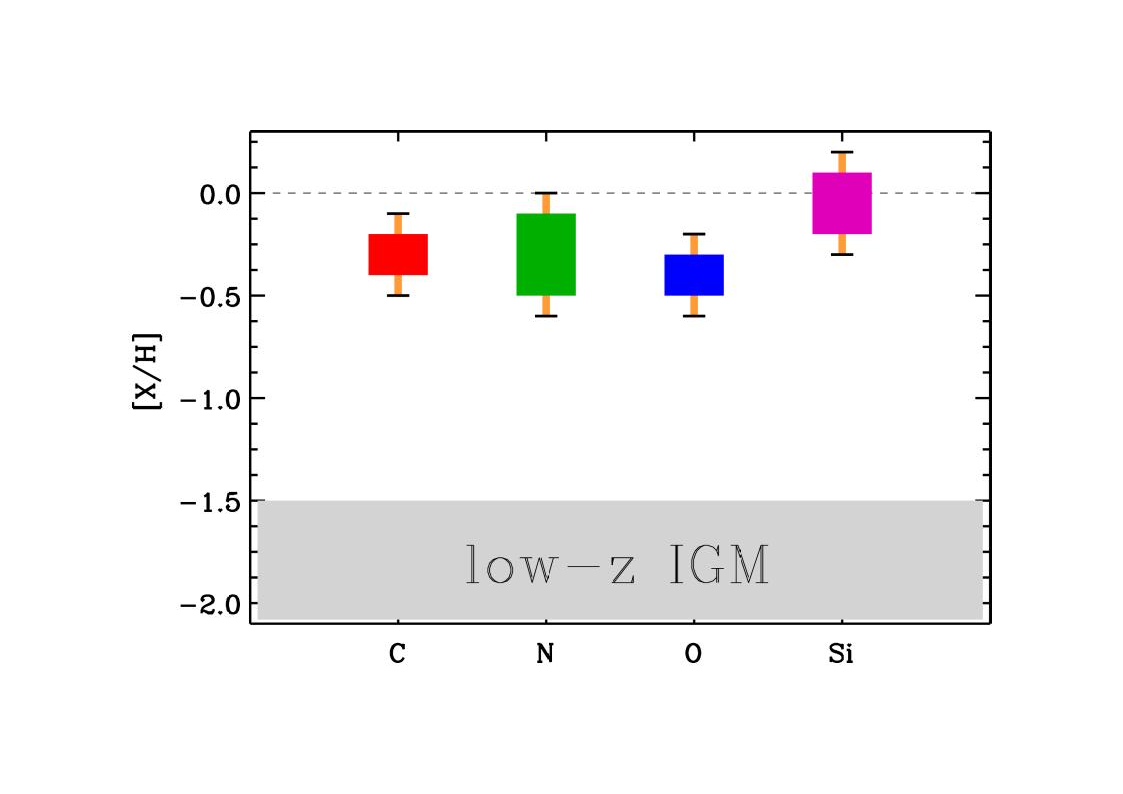}
         \end{subfigure}%
     \begin{subfigure}
         \centering
         \includegraphics[trim=2cm 2cm 2cm 2cm,scale=0.35]{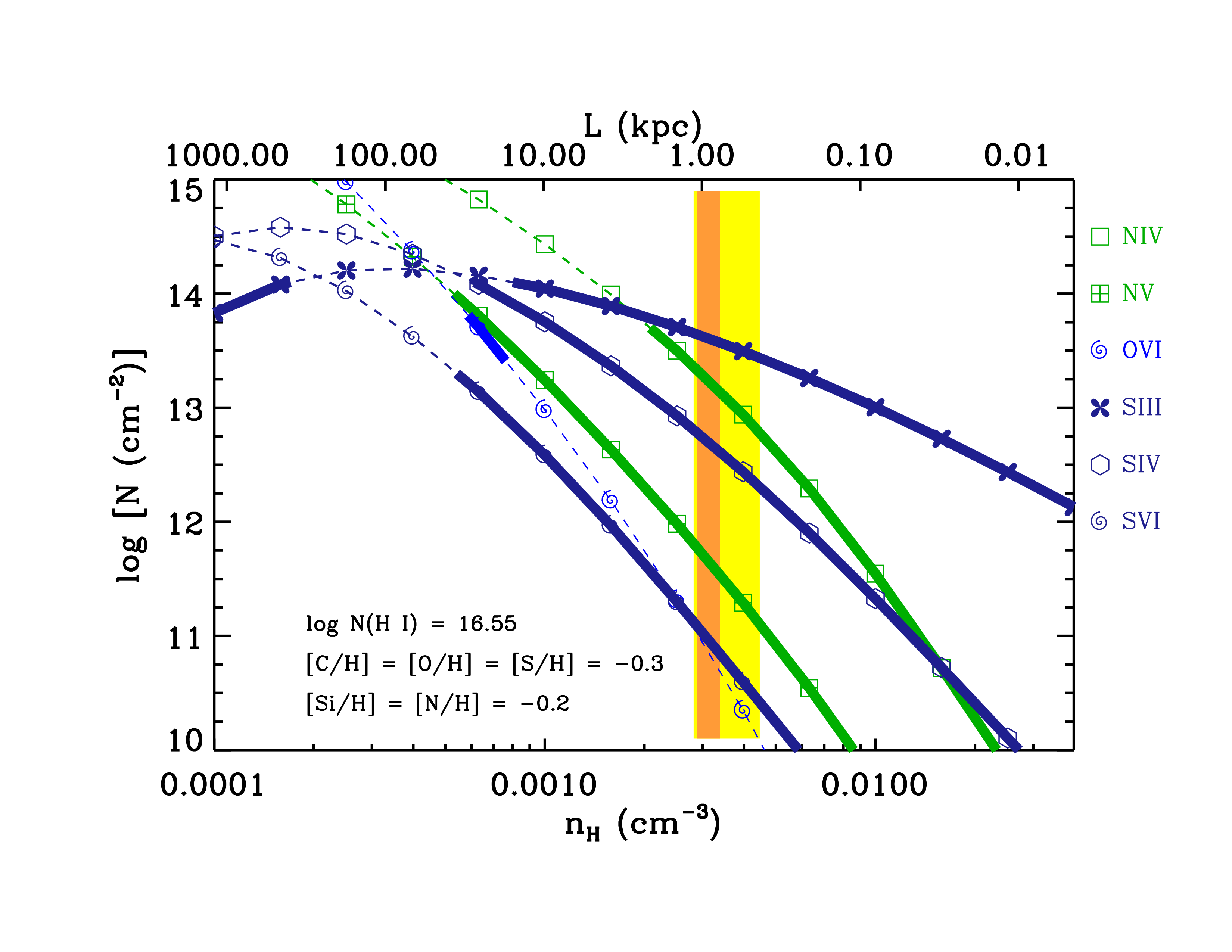}
         \end{subfigure}
     \caption{Photoionization equilibrium model for the z$_{abs}$ $=$ 0.43968 absorber towards UVQS J2017-4516. The model assumes an abundance of -0.3 solar for carbon, oxygen, sulphur and -0.2 solar for silicon and nitrogen. On the \textit{top left} is shown the model predicted (\textit{thin} lines) and the observed (\textit{thick} lines) column density ratios of successive ionization stages of the same element. This ratio, independent of metallicity, constrains the gas phase density in the absorber to a narrow range of densities indicated by the \textit{yellow} stripe. On the \textit{right} panels are shown the model-predicted variation of column densities of various species (thin lines), along with their observed values (thick lines), plotted against gas density (n$_{H}$). The density range allowed by the column density ratios is represented by the yellow region, while the final solution density range obtained by changing the chemical abundances is depicted in \textit{orange}. This single phase is consistent with all ions expect {\OVI} which requires a higher ionization phase. The line of sight thickness for a given density is given by the \textit{top} X-axis. On the \textit{left bottom} panel are the range of abundances permitted by the photoionization models for the different elements. The error bars correspond to uncertainty in estimating chemical abundances, carried over from the uncertainty in the $N(\HI)$ measurement. For comparison, the carbon abundance for the low-redshift IGM has been shown as the grey region (upper limit obtained from Barlow \& Tytler \protect\hyperlink{Barlow1998}{1998}).}
     \label{fig:sys2}
\end{figure*}

The detection of successive ionization stages of the same element, viz. {\CII}, {\CIII}, {\NII}, {\NIII} and {\SiII}, {\SiIII}, allow us to conclude on the gas phase density independent of metallicity. Among these ions, the column density ratio of {\NII} to {\NIII} offers the most reliable constraint as both lines are unsaturated. Both {\CIII} and {\SiIII} are saturated as reflected in the uncertainty in column density we obtain from profile fitting. In Figure \ref{fig:sys2} (top panel), we show the ionic column density ratios predicted by Cloudy as a function of gas density. The observed $\log~[N(\NII)/N(\NIII)] = -0.55~{\pm}~0.20$ is true for gas densities in the narrow range of $n_{\H}$ $=$ $(3 - 5)~\times~10^{-3}$~{\cc}. This density range is also consistent with the upper limits of  $\log~[N(\CII)/N(\CIII)] \leq 0.1$ and $\log~[N(\SiII)/N(\SiIII)] \leq 0.3$ obtained by considering the lower and upper $1\sigma$ limits in the column density for the low and high ionization stages of C and Si. The abundance limits can be set from the true column densities of {\CII}, {\NII}, {\OII} and {\SiII} coming from their respective unsaturated lines. From the photoionization models, we find that these ions are underproduced at all densities for abundances of [C/H] $< -0.4$, [N/H] = [O/H] $< -0.5$ and [Si/H] $< -0.2$~dex. At the same time, for [N/H] $\geq 0$, the predicted {\NII} falls outside of the density range given by the {\NII} to {\NIII} column density ratio. Thus, the nitrogen abundance in the absorber is constrained to values within the range $-0.5 \leq$~[N/H]~$< 0$. For the other low and intermediate ions to also have an origin in the same gas phase, the abundances should be within $-0.4 \leq$~[C/H]~$< -0.2$, $-0.5 \leq$~[O/H]~$< 0.2$, and $-0.2 \leq$~[Si/H]~$< 0.2$~dex. These elemental abundance ranges are shown in Figure \ref{fig:sys2} (bottom panel). A single phase solution at $n_{\H} \sim 3 \times 10^{-3}$~{\cc}, that agrees with all low and intermediate ions is also shown in Figure \ref{fig:sys2} (right panels). Such a phase will have a total hydrogen column density of $N(\HI)$ $=$ $10^{18.77}$~{\cmsq},
%$\log$~[N(\H)  (cm$^{-2}$)] = 18.77
pressure of $p/k = 47.1$~K cm$^{-3}$, a line of sight thickness of $L = 0.6$~kpc and a photoionization equilibrium temperature of $T = 1.5 \times 10^{4}$ K, consistent with the upper limit obtained from the neutral hydrogen line width.

The single phase solution, interestingly under-predicts the {\OVI} by $\sim 3$ orders of magnitude. The {\OVI} possibly arises from a separate phase of much higher ionization, and at higher temperatures of $T \gtrsim 10^5$~K where collisional ionization becomes important, as seen in the case of {\OVI} absorbers in galaxy overdensity environments (e.g., Narayanan {\etal}\hyperlink{Narayanan2010}{2010}, Pachat {\etal}\hyperlink{Pachat2016}{2016}). The observed {\HI} absorption will be dominated by the low ionization gas where the neutral fraction is going to be higher compared to the {\OVI} phase. Without information on the {\HI} that is exclusively associated with the {\OVI}, it is not possible to generate a model for the higher ionization gas. 

%Assuming a spherical geometry with diameter equal to L, the average mass of the cloud comes out to be around to be M = 3.192 x 10$^{7}$ M$\odot$.

\subsection{The z$_{abs}$ $=$ 0.51484 absorber towards UVQS J2109-5042}

\begin{figure*}
     \centering
     \begin{subfigure}
         \centering
         \includegraphics[trim=2cm 2cm 1cm 2cm,scale=0.35]{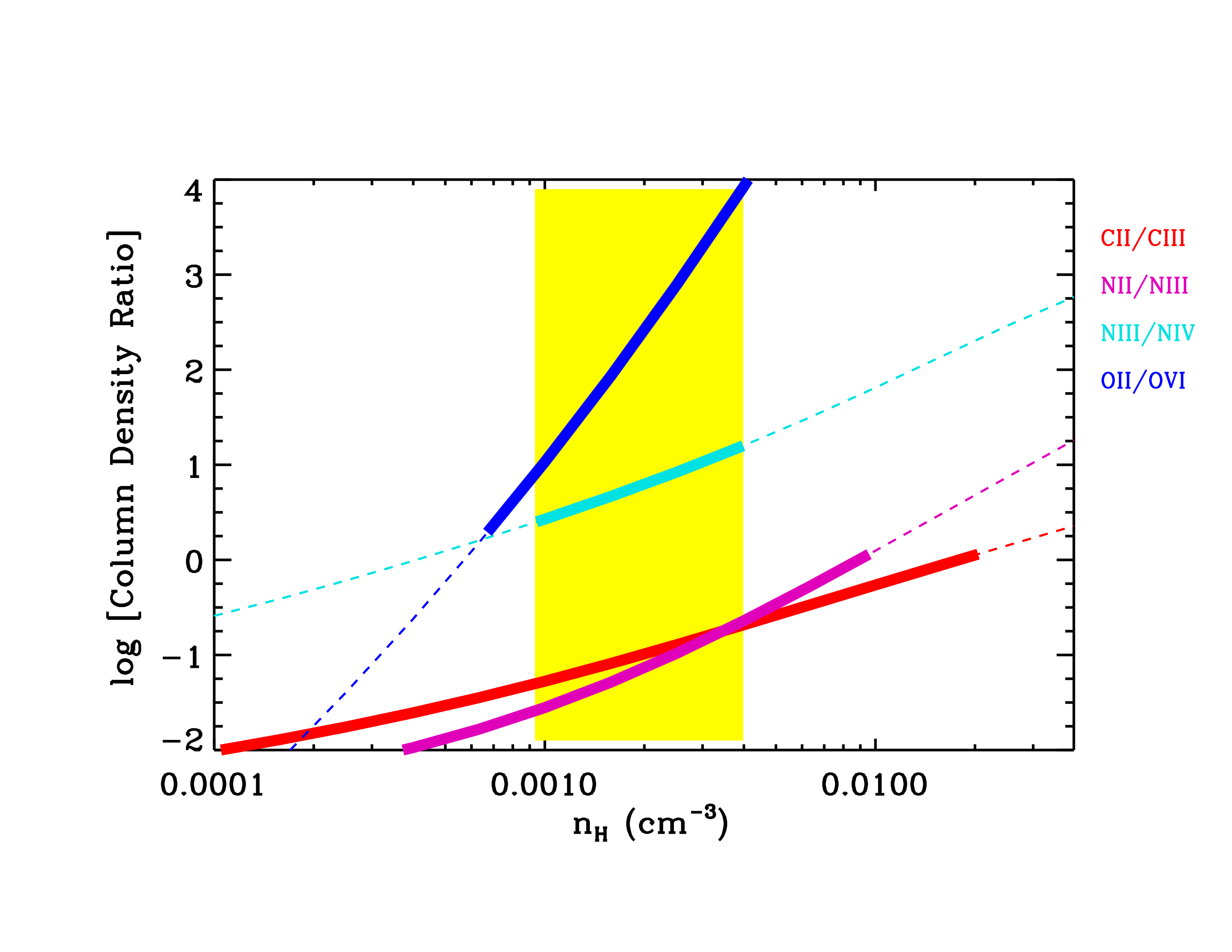}
     \end{subfigure}%
     \begin{subfigure}
         \centering
         \includegraphics[trim=2cm 2cm 2cm 2cm,scale=0.35]{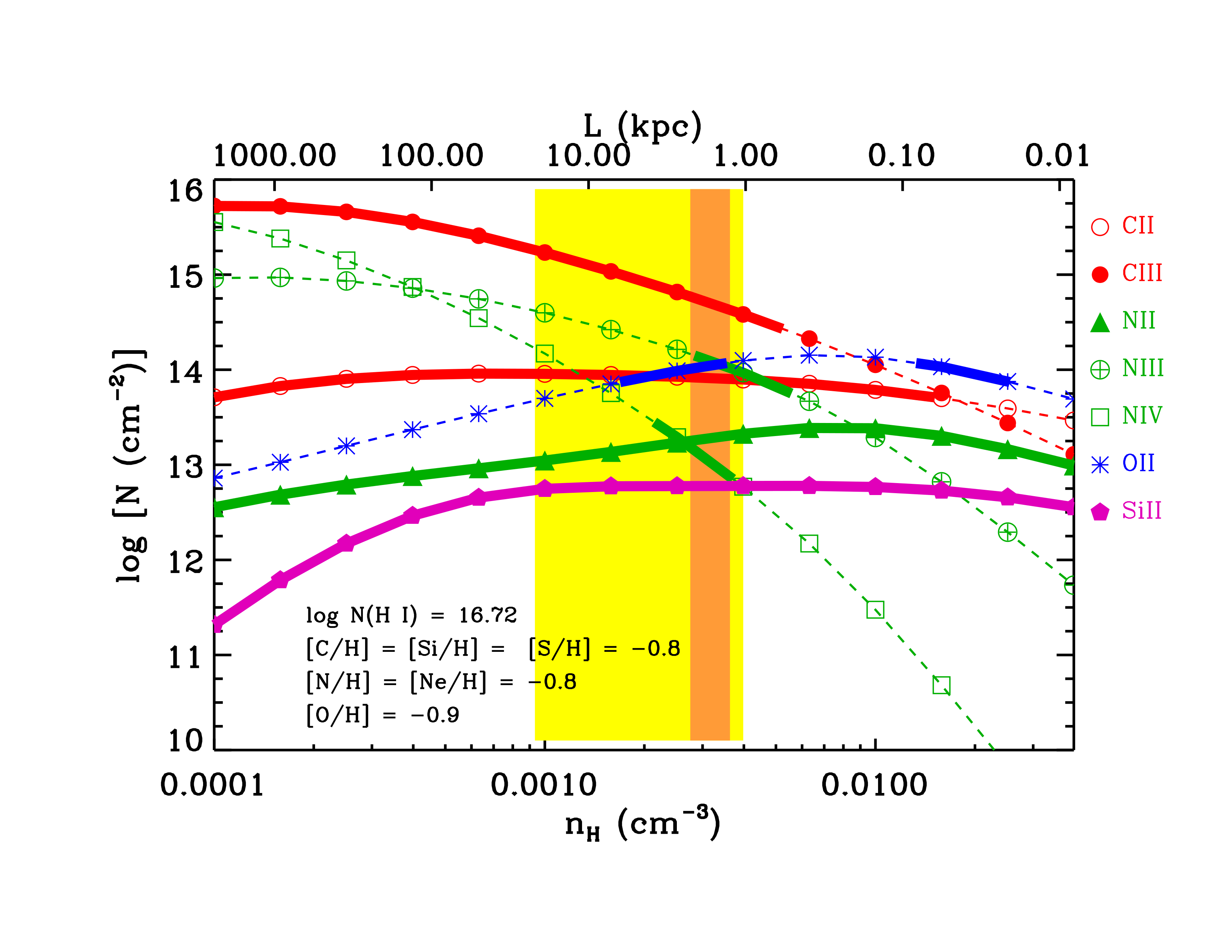}
     \end{subfigure}
     \begin{subfigure}
         \centering
         \includegraphics[trim=0cm 0cm 0.5cm 1cm,scale=0.35]{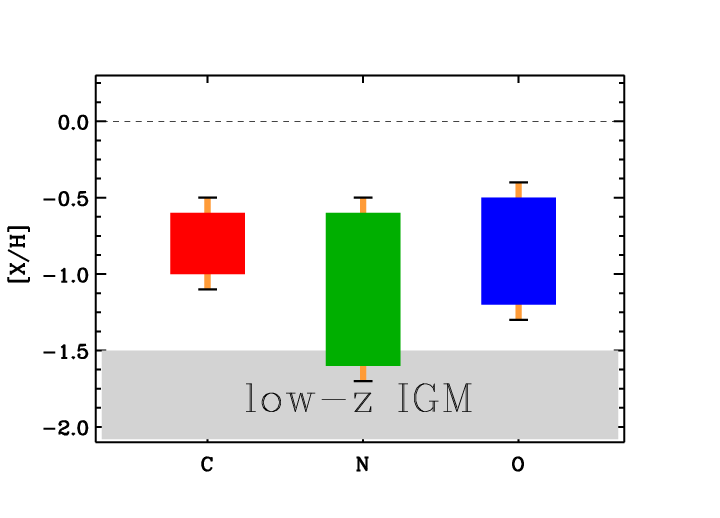}
     \end{subfigure}%
     \begin{subfigure}
         \centering
         \includegraphics[trim=2cm 2cm 2cm 2cm,scale=0.35]{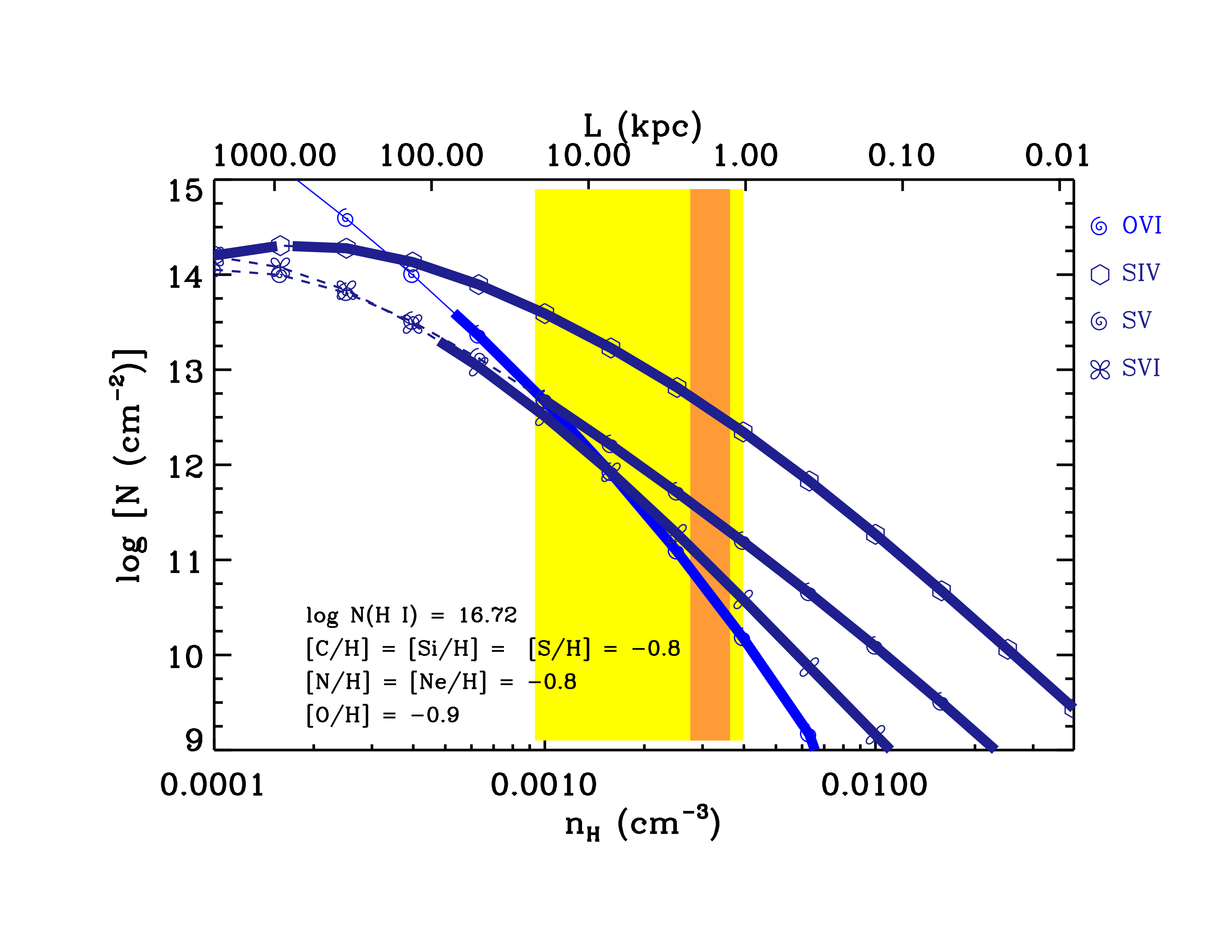}
     \end{subfigure}
     \caption{Photoionization equilibrium model for the z$_{abs}$ $=$ 0.51484 absorber towards UVQS J2109-5042, for the measured $N(\HI)$. The model assumes an abundance of -0.9 solar for oxygen and -0.8 solar for other elements. On the \textit{top left} is shown the model predicted (\textit{thin} lines) and the observed (\textit{thick} lines) column density ratios of successive ionization stages of the same element. This ratio, independent of metallicity, constrains the gas phase density in the absorber to a narrow range of densities indicated by the \textit{yellow} stripe. On the \textit{right} panels are shown the model-predicted variation of column densities of various species (thin lines), along with their observed values (thick lines), plotted against gas density (n$_{H}$). The density range allowed by the column density ratios is represented by the yellow region, while the final solution density range obtained by changing the chemical abundances is depicted in \textit{orange}. This single phase is consistent with all ions. The line of sight thickness for a given density is given by the \textit{top} X-axis. On the \textit{left bottom} panel are the range of abundances permitted by the photoionization models for the different elements. The error bars correspond to uncertainty in estimating chemical abundances, carried over from the uncertainty in the $N(\HI)$ measurement. For comparison, the carbon abundance for the low-redshift IGM has been shown as the grey region (upper limit obtained from Barlow \& Tytler \protect\hyperlink{Barlow1998}{1998}).}
     \label{fig:sys3}
\end{figure*}

The column density for {\CIII} in this absorber is taken as a lower limit to account for saturation, while the column densities of {\CII}, {\NIII}, {\NIV} and {\OII} are taken as measurements. The remaining metal lines are all non-detections, and provide useful upper limits on the column densities. The model-predicted variation of the column density ratios of successive ionization stages of the same element were used to establish the density. From this analysis, we identify n$_{H}$ = $(0.9 - 3.9)$ $\times$ 10$^{-3}$~{\cc} as a range within which all observed column \change{density} ratios can be simultaneously recovered, as shown in Figure \ref{fig:sys3}. An estimate for the metallicity can be arrived at by varying the chemical abundances of the metals to match their observed column densities within this density range given by the column density ratios. 

From the models, it was observed that $N(\CII)$ is underproduced at any density if [C/H] $< -1.0$.  [C/H] also has to be $< -0.5$ to recover the observed $N(\CII)$ within the acceptable range for density. The models reveal that the nitrogen abundance [N/H] must lie between $-1.7$ and $-0.5$ so that the observed  $N(\NIII)$ can be explained within the solution density range. The observed $N(\OII)$  cannot be recovered at any density if [O/H] $< -1.3$, while at the same time, adopting [O/H] $> -0.5$ will require a density that is outside the acceptable range given by the column density ratios. The resultant abundance ranges are shown in Figure \ref{fig:sys3} (left bottom panel). 
With these constraints, a single-phase solution that recovers the measured column densities was determined by adopting [X/H] = -0.8 for all the elements, except oxygen, for which [O/H] $= -0.9$ dex was taken. The difference is within the metallicity uncertainty of ${\pm}~0.15$~dex coming from the uncertainty in the {\HI} column density. This photoionization model solution is shown in Figure \ref{fig:sys3} (right panels), which is also consistent with the non-detections of {\SiII}, {\SIV}, {\SV}, {\SVI} and {\OVI}. This single phase solution predicts an average density of n$_{\H}$ = 3.2 $\times$ 10$^{-3}$~{\cc}, total hydrogen column density of $N(\HI)$ $=$ $10^{19.05}$~{\cmsq},
%$\log~[N(\H),~{\cmsq}] = 19.05$
pressure of $p/k = 42.37$~K~cm$^{-3}$ and line of sight thickness of $L = 1.2$~kpc. The solution also suggests a photoionization equilibrium temperature of $T = 1.34 \times 10^{4}$ K, in agreement with the prediction from the absorption line widths.

%Assuming a spherical geometry with the line of sight thickness as diameter, the average mass of the cloud is M = 6.29 x 10$^{4}$ M$\odot$.

\section{Discussion On the Origin of Absorbers}
\label{section:discussion}

In this section, we discuss possible origins of the three absorbers associated with three SZ-selected galaxy clusters at $z \sim 0.4 - 0.5$. The line strengths, gas densities, temperatures and intermediate to near-solar metallicity measurements of our absorbers serve as useful diagnostics on their astrophysical \change{origins}.

The absorbers we are studying could either be tracing diffuse cool-warm intracluster gas in the cluster outskirts, or clouds residing in the CGM of galaxies that are members of the clusters. The covering fraction of neutral gas in the CGM of cluster galaxies is estimated to be $18 - 25$\% for {\HI} with $W_r(\mathrm{Ly}\alpha) > 30$~m{\AA}, which is a factor of four lower compared to galaxies in the field (Yoon \& Putman \hyperlink{Yoon2013}{2013}, Burchett {\etal}\hyperlink{Burchett2018}{2018}). The most likely reason for this decline is the stripping of circumgalactic {\HI} from cluster galaxies by ram pressures from the ICM (Butsky {\etal}\hyperlink{Butsky2019}{2019}), as well as strong tidal forces that manifest during frequent galaxy interactions. The gas thus displaced can be found out to hundreds of kpc from cluster member galaxies (Boselli {\etal}\hyperlink{Boselli2016}{2016}, Gavazzi {\etal}\hyperlink{Gavazzi2018}{2018}), beyond the virial halos of low-mass galaxies, in the broader intracluster environment.

The metallicities for Lyman limit and partial Lyman limit systems at $z < 1$ display a bimodal distribution, peaking at $-1.8$~dex and $-0.3$~dex respectively (Lehner {\etal}\hyperlink{Lehner2013}{2013}), suggesting the presence of two different populations of high column density clouds at low redshifts, with possibly different origins. Lehner {\etal}(\hyperlink{Lehner2016}{2016}) \change{proposed} that the metal-rich population is most likely probing gas displaced from galaxies such as in outflows, tidal streams, or ram-pressure stripping, whereas the metal-poor population could be accretion from the intergalactic medium (IGM) (Keres {\etal}\hyperlink{Keres2005}{2005}, \hyperlink{Keres2009}{2009}, Dekel \& Birnboim \hyperlink{Dekel2006}{2006}, Brooks {\etal}\hyperlink{Brooks2009}{2009}, Dekel {\etal}\hyperlink{Dekel2009}{2009}, Cooper {\etal}\hyperlink{Cooper2016}{2016}, Zinger {\etal}\hyperlink{Zinger2016}{2016}).  Viewed in this context, the partial-Lyman limit column density, coupled with [C/H] $\sim -0.3$ estimated for the $z = 0.43968$ absorber, makes a case for chemically enriched gas originating from galaxies. Such an interpretation is also consistent with recent simulations of gas flows in galaxy clusters (Emerick {\etal}\hyperlink{Emerick2015}{2015}, Butsky {\etal}\hyperlink{Butsky2019}{2019}). These simulations reveal that {\HI} clouds with $N(\HI) \gtrsim 10^{14}$~{\cmsq} are often at moderate to solar metallicities, regardless of where they are positioned in the cluster environment. Ram-pressure stripping of chemically enriched CGM of cluster galaxies should create an extended  distribution of metal-rich gas that follows the large-scale galaxy distribution in and around clusters. 
 
 These simulations also show low column density absorbers with $N(\HI) \lesssim 10^{14}$~{\cmsq} as preferentially tracing cold-flow accretion from the IGM. Such infalling gas should also have low metallicities (e.g. Fumagalli {\etal}\hyperlink{Fumagalli2011}{2011}, Hafen {\etal}\hyperlink{Hafen2017}{2017}). In massive clusters, a scenario that is more relevant than cold-flow accretion from the IGM is that of warm penetrating streams from the cosmic web with T $>$ 10$^{5}$ K (Zinger {\etal}\hyperlink{Zinger2016}{2016}), which are also expected to be relatively metal poor. However, the areas subtended by these warm gas streams in clusters are quite small and therefore the probability of intercepting such flows in pencil-beam sightline observations remains low. Thus, flows of pristine gas streams from the IGM are unlikely to be the primary sources of high neutral column density, near-solar metallicity absorbers.

It is interesting to draw a comparison between the two partial Lyman-limit absorbers in our sample and the population of weak {\MgII} absorbers. Both absorbers have low ionization line strengths, primarily {\CII} and {\SiII}, consistent with weak {\MgII} systems (Narayanan {\etal}\hyperlink{Narayanan2005}{2005}). In the COS-Weak survey, Muzahid {\etal}(\hyperlink{Muzahid2018}{2018}) found as many as $80$\% of the weak absorbers in their low redshift ($z < 0.3$) sample to be residing in galaxy over-density regions, possibly galaxy groups, though only a small fraction of them ($14$\%) have a $> L^*$ galaxy within $50$~kpc of projected separation. The absence of close-by luminous star-forming galaxies (also reported by Churchill \& Le Brun \hyperlink{Churchill1998}{1998}, Churchill, Kacprzak \& Steidel \hyperlink{Churchill2005}{2005} and Milutinovi\'c {\etal}\hyperlink{Milutinovic2006}{2006}) is an important evidence for the origin of these weak absorbers, given the near-solar and higher metallicities generally inferred for them (Rigby {\etal}\hyperlink{Rigby2002}{2002}, Lynch \& Charlton \hyperlink{Lynch2007}{2007}, Misawa, Charlton \& Narayanan \hyperlink{Misawa2008}{2008}, Narayanan {\etal}\hyperlink{Narayanan2008}{2008}). It has been suggested that some substantial fraction of the weak absorbers could be tracing pockets of metal-rich gas displaced from galaxies in correlated supernova events, AGN winds or tidal stripping (Narayanan {\etal}\hyperlink{Narayanan2008}{2008}, Muzahid {\etal}\hyperlink{Muzahid2018}{2018}). The weak {\MgII} absorber analogues in our sample are also compatible with a similar origin in gas that is displaced from galaxies, and possibly assimilated into the ICM.

\change{An important alternative could be the origin of these absorbers in the larger intracluster volume itself. X-ray line measurements of metal abundances of galaxy clusters in the nearby universe have found metallicities to be uniformly high ([Fe/H] $\sim -0.5$) at different locations in the ICM even beyond the cluster radius of $\sim r_{200}$ (Werner {\etal}\hyperlink{Werner2013}{2013}, Urban {\etal}\hyperlink{Urban2017}{2017}), with the relative metal abundance ratios similar to solar (Simionescu {\etal}\hyperlink{Simionescu2015}{2015}). This spatial homogeneity in metallicity is attributed to an early enrichment of the IGM around large scale structures at a stage when clusters were still forming. The dispersion of metals into the cluster volume and the neighboring IGM environment must have been through AGN feedback as well as galactic winds from the enhanced rates of core-collapse and Type Ia supernovae operating on different timescales following the peak in the global star formation rate \dn{at} $z \sim 2 - 3$ (Fabjan {\etal}\hyperlink{Fabjan2010}{2010}, Werner {\etal}\hyperlink{Werner2013}{2013}, Biffi {\etal}\hyperlink{Biffi2018}{2018}). The cluster absorbers we study \dn{here} could be gas that has condensed out due to local thermal instabilities in this metal-enriched ICM, as \dn{proposed} for the origin of multiphase structures in the ICM by numerical models (\dn{e.g.,} McCourt {\etal}\hyperlink{McCourt2012}{2012}, Sharma {\etal}\hyperlink{Sharma2012}{2012}, McCourt {\etal}\hyperlink{McCourt2018}{2018}). Also, in a recent study, Mandelker {\etal}(\hyperlink{Mandelker2019}{2019}) have shown that thermal instability caused by strong shocks due to mergers of cosmic filaments lead to the formation of cool ($\sim$ 10$^{4}$~K) clouds with moderate densities ($\sim$ 10$^{-3}$ cm$^{-3}$) in regions far away ($\sim$ Mpc) from massive halos ($\sim$ 5 $\times$ 10$^{12}$ M$_{\odot}$) at $z=2$. These pristine ($<$ 10$^{-3}$ Z$_{\odot}$)  cool clouds in cosmic sheets, when viewed face-on, give rise to a significant covering fraction for Lyman Limit Systems.}

\change{If our absorbers indeed reside within the ICM in cluster outskirts, we can use the derived absorber densities and temperatures (reported in Table~\ref{table:summary}) to examine their thermal pressure balance with the surrounding ICM. The ICM pressure at the respective clustocentric projected radii can be estimated from the universal pressure profiles of Arnaud {\etal}(\hyperlink{Arnaud2010}{2010}). The computed ambient ICM pressure turns out to be comparable to the derived gas pressure in the case of the $z_{abs} = 0.43968$ absorber, and consistent with the lower limit on the gas pressure in the case of the $z_{abs} = 0.43737$ absorber, confirming that these two absorption systems are likely to have reached a state of thermal pressure equilibrium with the surrounding ICM. However, for the $z_{abs} = 0.51484$ absorber, which has a comparatively smaller clustocentric impact parameter, the estimated ICM pressure turns out to be 1.5 dex higher than the cloud pressure derived from photoionization models. This absorbing cloud is therefore likely to still be contracting under the influence of the ambient ICM pressure. Such a higher density cool cloud moving through a lower density hot ambient medium will experience Rayleigh-Taylor/Kelvin-Helmholtz instabilities which will eventually destroy the cloud (Klein, McKee \& Colella \hyperlink{Klein1994}{1994}, McCourt {\etal}\hyperlink{McCourt2015}{2015}).}

\begin{table*}\centering
\setlength{\tabcolsep}{5pt}
\renewcommand{\arraystretch}{1.5}
	\begin{tabular}{ccccccccc}
			\hline
			\centering
			 QSO & z$_{abs}$ & log N(HI) & n$_{\H}$ & log N$_{\H}$ & $p/k$ & T & L & [C/H]\\
			&  &  & (cm$^{-3}$) &  & (K cm$^{-3}$)  & (K) & (kpc) &  \\ \hline
			 \\
			UVQS J0040-5057  & 0.43737 & 18.63 $\pm$ 0.07 & $\gtrsim$ $5 \times 10^{-4}$ & $\lesssim$ 20.9
			 & $\gtrsim$ 8.6  & $ \lesssim 1.7 \times 10^{4}$ & $\lesssim$ 492.8 & $\geq$ -0.9 \\
		    \\

			UVQS J2017-4516 & 0.43968 & 16.55 $\pm$ 0.02 & $\sim$ 3 $\times$ $10^{-3}$ & $\sim$ 18.8 & $\sim 43.7$ & $\sim 1.5$ $\times$ $10^{4}$ & $\sim 0.9$ & $-0.35~{\pm}~0.10$ \\ 
			\\
			UVQS J2109-5042  & 0.51484 & 16.72 $\pm$ 0.05 & $\sim (0.9 - 3.9) \times 10^{-3}$ & $\sim [19.0,19.7]$ & $\sim [18.0, 60.9]$ & $\sim 1.7 \times 10^{4}$ & $\sim$ [0.9,18.4] & [-1.0,-0.6] \\\\ \hline
	\end{tabular}
	\caption{Summary of phase solution results from Photoionization Modelling of the three absorbers. The first column indicates the QSO along the line-of-sight of the absorber. Successive columns correspond to the absorber redshift (z$_{abs}$), the logarithm of neutral hydrogen column density ($\log N(\HI)$), the logarithm of total hydrogen column density ($\log N_{\H}$) measured from Lyman series lines, the solution phase gas density (n$_{\H}$), gas pressure ($p$) normalized by $k$, photoionization equilibrium temperature (T) and the path length (L) of the absorber along the sightline, indicating the size of the absorber in kpc. The final column indicates the obtained abundance of carbon ([C/H]) in each absorber.}
	\label{table:summary}
\end{table*}

Although the discussion so far pertains to a scenario in which the three absorbers in our sample are tracing metal-rich gas associated with the diffuse cool-warm ICM, another possibility is for these absorbers to be tracing chemically enriched gas within the halos of massive elliptical galaxies that are prominent members of clusters. Recent quasar absorption line surveys targeted at Luminous Red Galaxies (LRGs) at $z < 1$ have found chemically enriched gas of high column densities as prevalent in the CGM of those galaxies (e.g., Gauthier {\etal}\hyperlink{Gauthier2009}{2009}, Lundgren {\etal}\hyperlink{Lundgren2009}{2009}, Bowen \& Chelouche \hyperlink{Bowen2011}{2011}, Gauthier \& Chen \hyperlink{Gauthier2011}{2011}, Thom {\etal}\hyperlink{Thom2012}{2012}). High incidence of large $N(\HI)$ was found by Chen {\etal}(\hyperlink{Chen2018}{2018}) in LRG halos, with a covering fraction of $44$\% for Lyman limit absorbers at $d < 160$~kpc impact parameters. The intermediate ionization gas traced by {\CIII} and {\SiIII} has a $75$\% covering fraction at similar impact parameters, whereas {\OVI} is found to be not so widespread, with a covering fraction of only $18$\%. Similar estimates for {\HI} covering fraction are also obtained in the QSO-LRG absorption line surveys of Berg {\etal}(\hyperlink{Berg2018}{2018}), who also find a high covering fraction of high {\HI} column density absorbers ($N(\HI) > 10^{16}$~{\cmsq})
%who also find an absence of low {\HI} column density absorbers ($N(\HI) < 10^{16}$~{\cmsq}) 
in massive halos with $M_{*} > 10^{11.3}$~M$_{\odot}$ indicating that LRGs are a plausible candidate for optically thick {\HI} absorbers. Seen in {\HI} and intermediate ionization lines, massive quiescent galaxy halos and star-forming halos turn out to be indistinguishable with both possessing a high covering fraction of cool low ionization gas (Werk {\etal}\hyperlink{Werk2013}{2013}, Chen {\etal}\hyperlink{Chen2018}{2018}). The significantly lower incidence of {\OVI} is the factor that differentiates passive halos from star forming ones (Tumlinson {\etal}\hyperlink{Tumlinson2011}{2011}, Werk {\etal}\hyperlink{Werk2013}{2013}). Berg {\etal}(\hyperlink{Berg2018}{2018}) interpret the metal rich Lyman and partial Lyman limit LRG-CGM absorbers as cool clouds born out of thermal instabilities in the hot corona of massive elliptical systems. The absence of {\OVI} in such absorbers could be due to a lack of gas with densities that are low enough ($n_{\H} \lesssim 10^{-5}$~{\cmsq}) to produce {\OVI} in large amounts through photoionization. With no clear evidence for on-going star formation, except in a minority of LRGs, the abundance of metals in vast majority of the LRG halos have to be from prior episodes of star formation, or past central AGN activity.

The absorbers in our study show interesting resemblances to LRG-CGM absorbers in terms of absorption line properties, %prompting us to emphasize upon the possibility of 
indicating a possibility that they could be gas clouds associated with the CGM of LRGs in the respective clusters. Although massive clusters are expected to have several luminous elliptical galaxies in them, our sightlines are probing regions far away from the cluster centre. Hoshino {\etal}(\hyperlink{Hoshino2015}{2015}) reported that in a large number of clusters the brightest LRG is not the central galaxy of the cluster, but they also observed that the radial distribution of non-central LRGs in clusters is substantially truncated at the outskirts. This indicates a possibly low compound probability for our three separate lines of sight to be probing in each case the CGM of LRGs in cluster outskirts. Nevertheless, dedicated deep galaxy surveys of these fields are essential to firmly establish whether or not our absorbers indeed belong to cluster galaxies.%, in particular LRGs. 

\section{Conclusions}\label{section:conclusions}

We have undertaken a study of the properties of {\HI} and metal lines in three strong {\HI} absorbers at redshifts $z =$ $0.43737$, $0.43968$, \& $~0.51484$, associated with three SZ-selected galaxy clusters, \change{the properties of which are summarized in Table \ref{table:qsocluster}.  The clustocentric impact parameters indicate that the absorbers are located away from the hot central X-ray emitting regions of the clusters. They show substantial lines of sight velocities of $-2600$~{\kms}, $-2100$~{\kms} and $9000$~{\kms} with respect to the corresponding redshifts of the clusters, which are $z_{cl} =$ $0.45$, $0.45$ \& $0.47$ (Bleem {\etal}\hyperlink{Bleem2015}{2015})}. \change{These cluster redshifts are photometric and carry large errors of $\sim$ 0.04, which bring the 
velocity offsets of the absorbers well within the uncertainty associated with each cluster redshift (|$\Delta$z/(1 + z)| $\approx$ 0.03). From the redshift evolution of low-z {\HI} absorbers, Muzahid {\etal}(\hyperlink{Muzahid2017}{2017}) computed the compound probability of random occurrence of the three absorbers so close to the cluster redshifts to be substantially low ($<$ 0.02\%), 
%strongly 
\dn{indicating} that the absorbers are indeed associated with the corresponding clusters}. The key results and conclusions from our analysis of these absorbers are listed in this section.

\begin{enumerate} 

\item The partial Lyman limit and Lyman limit column densities make these the highest {\HI} column density absorbers known thus far in clusters. The widths of {\HI} and metal lines indicate gas temperatures of $T \sim 10^4$~K. Photoionization models produce self-consistent single phase models with gas densities of $n_{\H} \sim 10^{-3}$~{\cc}, temperatures in agreement with the prediction from line widths, and metallicities in the range of one-tenth solar to near solar.  The ionization modelling results are summarized in Table~\ref{table:summary}. 

%\item From the derived densities and temperatures reported in Table~\ref{table:summary}, it is evident that there could be pressure imbalance between the absorbers and the surrounding ICM, assuming a temperature of $10^7$ K and density of 10$^{-4}$ cm$^{-3}$ for the latter. In such a scenario, the clouds can be confined by their self-gravity only if the absorber size is greater than the Jeans scale, as given in Schaye (\hyperlink{Schaye2001}{2001}) and Schaye (\hyperlink{Schaye2007}{2007}). For our clouds, the absorber size is well-constrained only for the $z_{abs} = 0.43968$ absorber, with a value of $\sim$ $0.9$~kpc, which is much less than the corresponding Jeans scale of $\sim 5.8$~kpc, obtained by adopting a gas mass fraction equal to the universal gas mass fraction of one-sixth. For CGM or intracluster clouds, the gas mass fraction and the corresponding Jeans size will be higher. Thus, at least in the case of this absorber, we could be looking at a transient cloud. 
%The absorber sizes of the remaining clouds are not constrained rigorously enough for us to comment on the

\item We report a strong constraint on the near-solar metallicity of the $z = 0.43968$ absorber, indicating a possible origin via the stripping of metal-enriched gas from the CGM of cluster galaxies. The [C/H] = $-0.35~{\pm}~0.10$ for the $z = 0.43968$ absorber is accurately estimated and is higher than what is measured generally for the low-$z$ IGM ([X/H] $< -1.0$; e.g., Barlow \& Tytler \hyperlink{Barlow1998}{1998}, Danforth {\etal}\hyperlink{Danforth2005}{2005}), but is comparable with the intermediate metallicities of the hot intracluster medium in galaxy clusters (Baldi {\etal}\hyperlink{Baldi2007}{2007}, Balestra {\etal}\change{\hyperlink{Balestra2007}{2007}}). The partial-Lyman limit column density, coupled with [C/H] $\sim -0.3$ estimated for the $z = 0.43968$ absorber, make a case for chemically enriched gas removed from galaxies, rather than pristine gas streaming in from the IGM.

\item For the other two absorbers, metallicities are not as robustly constrained. Nonetheless, photoionization models based on the available absorption lines suggest a lower bound of [X/H] $\gtrsim -0.9$ for both absorbers, characteristic of the high metallicity branch of the population of Lyman limit absorbers in the low redshift universe. Hence, the high neutral gas column densities and metallicities of the other two absorbers also point at an origin similar to the $z = 0.43968$ absorber. 

\item \change{An alternative is for the absorbers to be tracing cool ($T \sim 10^4$~K)  
%condensations  %born out of 
\dn{gas condensing out of the ICM itself, via thermal instabilities}
(McCourt {\etal}\hyperlink{McCourt2012}{2012}, Sharma {\etal}\hyperlink{Sharma2012}{2012}). 
\dn{In this scenario, the relatively high metal abundances we derived are consistent with the uniformly high metallicities in cluster outskirts inferred by recent X-ray observations, and interpreted to be arising from an earlier epoch of supernova and AGN feedback (Werner {\etal}\hyperlink{Werner2013}{2013}, Simionescu {\etal}\hyperlink{Simionescu2015}{2015} and Urban {\etal}\hyperlink{Urban2017}{2017}).}}

%In such a scenario, the relatively high metal abundances we \dn{derived} can be justified by recent observational 
%and theoretical 
%indications of uniformly high metallicities even in the outskirts of the ICM of nearby galaxy clusters, arising from an earlier epoch of supernova and AGN feedback (Werner {\etal}\hyperlink{Werner2013}{2013}, Simionescu {\etal}\hyperlink{Simionescu2015}{2015} and Urban {\etal}\hyperlink{Urban2017}{2017}).}  

\item The absorption line properties of the absorbers in our sample are also remarkably similar to the CGM of LRGs. Our absorbers have strong {\CII}, {\CIII}, {\NIII} and {\SiIII} lines, consistent with the high covering fractions of these species in LRG-CGM absorbers. Also, {\OVI} is a non-detection in two of our cases and a marginal $3\sigma$ detection in the third one, consistent with the lack of {\OVI} in LRG-CGM absorbers. The temperatures and densities obtained from photoionization modelling agree with the generic temperatures ($T \sim 2 \times 10^4$~K) and densities ($n_{\H} \sim (0.2 - 1) \times 10^{-3}$~{\cc}) seen for the population of CGM absorbers around LRG galaxies (Zahedy {\etal}\hyperlink{Zahedy2019}{2019}). The sub-solar metallicities we obtain are also consistent with what Zahedy {\etal}(\hyperlink{Zahedy2019}{2019}) estimate for LRG circumgalactic clouds, with $50$\% of the absorbers in their sample having [X/H] $> -1.0$~dex. However, the compound probability of having all three sightlines intersecting the CGM of LRGs is expected to be very small, especially in cluster outskirts. Therefore, although our absorbers evidently appear to be similar to the CGM of LRGs, this does not necessarily explain why we see such strong HI absorption in all three cases. Since we do not have any information about galaxies near these quasar sightlines, spectroscopic galaxy surveys in the fields around our absorbers are needed to better assess this scenario.

\item The absorbers in this study exhibit a notable absence of strong {\OVI} absorption. {\OVI} is a weak ($3\sigma$) detection compared to other ionization stages of oxygen in one of the absorbers, and is a non-detection in the remaining two. It is known that {\OVI} can be produced through collisional ionization in the conductive interface layers between relatively cool ($T \sim$ 10$^{4}$ K) gas and a hotter ($T \geq 10^6$~K) ambient medium such as the hot corona of a galaxy or the ICM. The {\OVI} detected in the $z = 0.43968$ absorber could have an origin in such an interface layer where the ionizations are dominated by electron - ion collisions at $T \gtrsim 10^5$~K. The lack of {\OVI} in the other two absorbers could be pointing at the absence of such a dense interface layer in these systems. 

\end{enumerate} 

%In summary, the Lyman limit and partial Lyman limit absorbers discussed in this work are either (a) tracing a phase of the ICM that is cooler than the hot X-ray emitting regions, with chemical abundances indicative of circumgalactic gas removed from cluster galaxies, 
%that belong to the cluster, 
%or (b) cool circumgalactic clouds embedded within the baryon rich halos of massive elliptical galaxies that coincide with cluster environments. Deep galaxy spectroscopic surveys of these fields are needed to distinguish between the possible astrophysical origins of these absorbers. 

To summarize, the Lyman limit and partial Lyman limit absorbers discussed in this work are most likely to be tracing a phase of the ICM that is cooler than the hot X-ray emitting regions, with chemical abundances indicative of \change{either circumgalactic gas removed from cluster galaxies, or early metal-enrichment in the ICM itself}. Generating a larger sample of such cluster absorbers through future observations with $HST/COS$ can add essential detail to our understanding of the multiphase gas properties in the ICM. 

%In addition, more number of targeted observations of lines of sight through galaxy overdensity regions harnessing the UV capabilities of $HST$, may improve the sample size of such cluster absorbers. 
%Establishing the chemical abundances in these absorbers using strong constraints set by the metal lines and modelling can help differentiate between accretions of pristine IGM onto galaxy clusters from the recycling of chemically processed gas belonging to the cluster itself. 

\section*{Acknowledgements}
\dn{The authors wish to thank the anonymous referee for the careful scrutiny of the manuscript and the valuable comments.} Support for this work was provided by SERB through grant number EMR/2017/002531 from the Department of Science \& Technology, Government of India. Based on observations made with the NASA/ESA Hubble Space Telescope, support for which was given by NASA through grant HST GO-14655 from the Space Telescope Science Institute. STScI is operated by the Association of Universities for Research in Astronomy, Inc. under NASA contract NAS 5-26555. This research has made use of the HSLA database, developed and maintained at STScI, Baltimore, USA.

\appendix

\section{Tables of Measurement \label{section:tables}}

\newcolumntype{Y}{>{\centering\arraybackslash}X}

\begin{table}\centering
\setlength{\tabcolsep}{1.5pt}
\renewcommand{\arraystretch}{1.5}
\caption{Line measurements for the  z$_{abs}$ $=$ 0.43737 absorber towards UVQS J0040-5057 with the successive columns indicating the corresponding equivalent width in the rest-frame of the absorber, the column density measured through the AOD method and Voigt profile fitting and the Doppler parameters obtained through profile fitting. The final column shows the velocity range over which the equivalent width and apparent column densities were integrated, or the centroid for the profile-fitted absorption components. \change{Note that the errors in column densities are likely to be underestimated as continuum placement uncertainties have not been accounted for.}}
\label{table1}
\begin{tabularx}{\linewidth}{@{}p{4em}YYYYY@{}}

\toprule
\multicolumn{1}{c}{Line} & \multicolumn{1}{c}{$\mathit{W_{r}}(m\mathit{A^{o}})$} &  \multicolumn{1}{c}{log[N ($\mathit{cm^{-2}}$)]} & \multicolumn{1}{c}{b(km/s)} &  \multicolumn{1}{c}{v (km/s)}  \\ \midrule

H I 1215 &   1763 $\pm$ 70 & $>$ 14  &  
& [-380,330]  \\

H I 1025 &  820 $\pm$ 14 & $>$ 15  &  & [-180,120] \\

H I 972 &  776 $\pm$ 24 & $>$ 15   &   & [-180,120]  \\

H I 949 &   723 $\pm$ 15 & $>$ 16    &   & [-180,120] \\

H I 937 &   694 $\pm$ 9 & $>$ 16    &  & [-180,120] \\

H I 930 &   666 $\pm$ 9 & $>$ 16   &  & [-180,120] \\

H I 926 &  637 $\pm$ 9 & $>$ 16   &  & [-180,120] \\

H I 923 &   641 $\pm$ 11 & $>$ 17   &  & [-180,120] \\

H I 920 &   659 $\pm$ 10  & $>$  17   &  & [-180,120] \\

H I 919 &   613 $\pm$ 11 & $>$  17   &  & [-180,120] \\

H I  & &  16.51 $\pm$  0.06 
& 16 $\pm$ 1 & -118 $\pm$ 4 \\

 &  &  18.56    $\pm$ 0.15  
&  23 $\pm$ 2  & -38 $\pm$ 3 \\

 &  &  18.50    $\pm$ 0.12  
& 14 $\pm$ 2 & 36 $\pm$ 4   \\

C II 1036 &   396 $\pm$   21  & 14.75 $\pm$ \change{0.09}  &  
 & [-170,100] \\

C II 1036 &  &  13.95 $\pm$   0.09  
 & 15 $\pm$ 6 & -118  $\pm$ 4  \\

&  &  15.03 $\pm$  0.11  & 23 $\pm$ 3 & -38  $\pm$    3\\ 

&  &  14.09 $\pm$  0.09   & 15 $\pm$ 8 & 36  $\pm$  4 \\

C III 977 &   644 $\pm$      25   & $>$ \change{14.3}  &  
 & [-170,100] \\

C III 977 &  &  14.04 $\pm$   0.22  
&  15 $\pm$ 3 & -115  $\pm$ 6\\

&  &  15.35 $\pm$  0.41   &  21 $\pm$ 3 &  -40  $\pm$    7\\ 

&  &  14.43 $\pm$  0.26   & 16 $\pm$ 3 &  35  $\pm$  8\\

N II 1083 &   281 $\pm$      23   &  14.63 $\pm$ \change{0.16}  &  
 & [-170,100]  \\

N II 1083 &  &  14.59 $\pm$   0.06   
&  23 $\pm$ 3 & -42  $\pm$ 3\\

N III 989 &  278 $\pm$     33   &  14.65 $\pm$ \change{0.19}  &  
 & [-170,100] \\

N III 989  &  &  14.44 $\pm$   0.09  
& 23 $\pm$ 3 & -44  $\pm$ 3 \\

N V 1238 &  $<$  \leavevmode\change{320}  & $<$  \leavevmode\change{14.2}  &  & [-170,100] \\

{\fontsize{7.6}{60}\selectfont O VI 1031} &  $<$  69  & $<$ \leavevmode\change{13.7} & & [-170,100]  \\

{\fontsize{7.6}{60}\selectfont O VI 1037} &  $<$   \leavevmode\change{74}  & $<$  \leavevmode\change{14.0}   &  
 & [-170,100]  \\

Si II 1193 &    245 $\pm$ 46  &   13.7 $\pm$ \change{0.31}   &  
 & [-170,100]  \\

Si II 1190 &  $<$   \leavevmode\change{143} & $<$  \leavevmode\change{13.6} &  
 & [-170,100]  \\

Si II 1020 &  $<$  \leavevmode\change{69} & $<$ \leavevmode\change{14.6}     &  
& [-170,100]   \\

Si II  &  & 13.58 $\pm$   0.08   
& 23 & -51  $\pm$ 3\\

S III 1012 &  $<$   \leavevmode\change{65} & $<$  \leavevmode\change{14.3}     &  
& [-170,100]  \\

S VI 944 &  $<$   44 & $<$ \leavevmode\change{13.4}      &  
 & [-170,100] \\

S VI 933 &  $<$  41 & $<$  \leavevmode\change{13.1}     &  
& [-170,100]   \\

{\fontsize{7.9}{60}\selectfont Fe II 1096} &  $<$   80  & $<$  \leavevmode\change{14.4} &  
 & [-170,100]  \\

{\fontsize{7.9}{60}\selectfont Fe II 1081} & $<$   79  & $<$  14.8 &  
 & [-170,100]  \\

\bottomrule
\end{tabularx}
\end{table}

\newcolumntype{Y}{>{\centering\arraybackslash}X}

\begin{table}\centering
\setlength{\tabcolsep}{2.5pt}
\renewcommand{\arraystretch}{1.5}
\caption{Line measurements for the z$_{abs}$ $=$ 0.43968 absorber towards UVQS J2017-4516 with the successive columns indicating the corresponding equivalent width in the rest-frame of the absorber, the column density measured through the AOD method and Voigt profile fitting and the Doppler parameters obtained through profile fitting. The final column shows the velocity range over which the equivalent width and apparent column densities were integrated, or the centroid for the profile-fitted absorption components. \change{Note that the errors in column densities are likely to be underestimated as continuum placement uncertainties have not been accounted for.}}
\label{table2}

\begin{tabularx}{\linewidth}{@{}p{5em}YYYYY@{}}

\toprule
\multicolumn{1}{c}{Line} & \multicolumn{1}{c}{$\mathit{W_{r}}(m\mathit{A^{o}})$} &  \multicolumn{1}{c}{log[N ($\mathit{cm^{-2}}$)]} & \multicolumn{1}{c}{b(km/s)} &  \multicolumn{1}{c}{v (km/s)}  \\ \midrule

H I 1215 &   521 $\pm$  29 & $>$ 14  &  
 & [-95,80]  \\

H I 1025 &   357 $\pm$      13 & $>$ 15  &  
& [-95,80] \\

H I 972 &  311 $\pm$      19 & $>$ 15   &  
& [-95,80]   \\

H I 937 &   236 $\pm$     10  & $>$ 16    &  
& [-95,80] \\

H I 930 &    240 $\pm$      10 & 16.11  $\pm$    0.03 &  & [-95,80]  \\

H I 923 &   197 $\pm$   10 & 16.32 $\pm$   0.03 &  
& [-95,80]  \\

H I 920 &   155 $\pm$     11   & 16.31 $\pm$     0.03 &  & [-95,80]\\

H I 919 &    124 $\pm$     10 & 16.38 $\pm$    0.03 & & [-95,80] \\

H I 918 &    134 $\pm$    11 & 16.47 $\pm$  0.03 &  
& [-95,80]  \\

H I 917 &   102 $\pm$ 10  &  16.44 $\pm$ 0.04   &  
& [-95,80] \\

H I  &     &  16.55 $\pm$   0.02    
& 23 $\pm$ 2 & -6 $\pm$ 1   \\

%H I  &     &  16.55 $\pm$   0.02    
%& 22 $\pm$ 1 & -6 $\pm$ 1   \\ % actual

%C II 903.62 &  144 $\pm$    11   & 14.22 $\pm$ 0.01  &  
%[-70,55] & & \\

C II 1036 &   122 $\pm$     14   & 14.15 $\pm$ 0.05  &  &  [-70,55] \\

C II 1036 &  &  14.19 $\pm$   0.09   
 & 23 $\pm$ 3 & -4  $\pm$ 3      \\

C III 977 &   244 $\pm$    18   & $>$ 13.9   &  
 & [-70,55]  \\

C III 977 &  &  14.49 $\pm$   0.32    
&  23 $\pm$ 3 & -3 $\pm$ 2 \\

%N II 1083 &   46 $\pm$    14  &  13.76 $\pm$ 0.02  &  
%[-35,35]* & & \\

N II 915 &    58 $\pm$  10   &  13.76 $\pm$ 0.05  &  
& [-60,25] \\

N II 915 &  &  13.70   $\pm$   0.12   
&  23 $\pm$ 3 & -2 $\pm$ 1 \\

N III 989 &  128 $\pm$     21  &  14.29 $\pm$ \change{0.13}  &
 & [-70,55] \\

N III 989 &  &  14.25 $\pm$   0.16     
&  23 $\pm$ 3 & 2 $\pm$ 4  \\

N IV 765 &  $<$ 146  &  $<$ \change{13.7} &  
& [-70,55]  \\

N V 1238 &  $<$  184  & $<$  \change{14.0}    &  
& [-70,55]  \\

O II 834 &  121 $\pm$   10  &    14.36 $\pm$     0.04 &  & [-70,55] \\

O II 834  &  &  14.49 $\pm$   0.09   
& 23 $\pm$ 3 & 0.2  \\

%O III 832 &  $<$ 210   &   $<$ 14.73    &  	 
%[-70,55]$^{3}$ & \\

%O IV 787 &  167 $\pm$       15   & $<$ 14.77 &  
%[-70,40]$^{4}$ & \\

O VI 1031 & 46 $\pm$ 15 & \leavevmode\change{13.61} $\pm$ 0.14    &  
& [-70,55]   \\

%O VI 1031 &   46  $\pm$ 15 & 13.68 $\pm$ 0.14    &  
%& [-70,55]   \\

%O VI 1037 &   $<$ 36   & $<$ 14.49    &  
%[-70,55]$^{5}$ &  \\

Ne VIII 770 &  $<$   109  & $<$ 14.3     &  
 & [-70,55]  \\

Ne VIII 780 &  $<$   69  & $<$  \change{14.4} &  
& [-70,55]   \\

Si II 1193 &  127 $\pm$    27  &  13.36 $\pm$  \change{0.19}  &  & [-70,55]   \\

Si II 1020 &  $<$   41  & $<$  14.4 &  
& [-70,55]   \\

Si II   &  &  13.35 $\pm$   0.17  
& 23 $\pm$ 3 & -5 $\pm$ 3 \\

Si III 1206 &  181 $\pm$ 26 & 13.2 $\pm$ \change{0.23} &  
& [-70,55]   \\

Si III 1206  &  & 13.33  $\pm$   0.15  
& 23 $\pm$ 3 & -5 $\pm$ 2  \\

S III 1012 &  $<$  40   & $<$  14.1     &  
& [-70,55]   \\

S IV 1062 &  $<$   47 & $<$  \change{14.1} &  
& [-70,55]   \\

S VI 944 &  $<$  33 & $<$  \change{13.3}     &  
& [-70,55]   \\

%S VI 933 &  $<$ 33    & $<$ 13.17  &  
%[-70,55] &  \\
\bottomrule
\end{tabularx}
\end{table}

\newcolumntype{Y}{>{\centering\arraybackslash}X}

\begin{table}\centering
\setlength{\tabcolsep}{2.5pt}
\renewcommand{\arraystretch}{1.5}
\caption{Line measurements for the z$_{abs}$ $=$ 0.51484 absorber towards UVQS J2109-5042 with the successive columns indicating the corresponding equivalent width in the rest-frame of the absorber, the column density measured through the AOD method and Voigt profile fitting and the Doppler parameters obtained through profile fitting. The final column shows the velocity range over which the equivalent width and apparent column densities were integrated, or the centroid for the profile-fitted absorption components. \change{Note that the column density errors are likely to be underestimated as continuum placement uncertainties have not been accounted for.}}
\label{table3}

\begin{tabularx}{\linewidth}{@{}p{5em}YYYYY@{}}

\toprule
\multicolumn{1}{c}{Line} & \multicolumn{1}{c}{$\mathit{W_{r}}(m\mathit{A^{o}})$} &  \multicolumn{1}{c}{log[N ($\mathit{cm^{-2}}$)]} & \multicolumn{1}{c}{b(km/s)} &  \multicolumn{1}{c}{v (km/s)}  \\ \midrule

H I 1025 &  \leavevmode\change{492} $\pm$    20 & $>$ 15  &  
 & [-155,90] \\

H I 972 &  \leavevmode\change{395} $\pm$  18 & $>$ 15   &  
 & [-155,90]  \\

H I 937 &  \leavevmode\change{302} $\pm$ 27 & $>$ 16    &  
 & [-155,90] \\

H I 930 &   \leavevmode\change{374} $\pm$   25 & $>$ 16   &  
 & [-155,90]  \\

H I 926 &  \leavevmode\change{345} $\pm$    25 & $>$ 16   &  
 & [-155,90] \\

H I 923 &  294 $\pm$   25 & $>$ 16  &  
 & [-155,90] \\

H I 920 &   \leavevmode\change{232} $\pm$   26  & $>$  16   &  
 & [-155,90] \\

H I 919 &  \leavevmode\change{223} $\pm$     26 & 16.72  $\pm$   \change{0.28}  &  
 & [-155,90] \\

H I 918 &   \leavevmode\change{157} $\pm$  27 & 16.74  $\pm$    \change{0.25} &  
 & [-155,90] \\ 

H I 917 &   \leavevmode\change{166} $\pm$    28 &  16.78  $\pm$     \change{0.18} &  
 & [-155,90] \\

H I &     & 16.72    $\pm$ 0.05    
& 30 $\pm$ 2  &  -20 $\pm$ 2 \\

C II 1036 &   \leavevmode\change{51} $\pm$    \change{20}   &  13.79  $\pm$   0.15  &  
 & [-80,45]  \\

C II 1036 &  &  13.91   $\pm$  0.11   & 14   $\pm$ 4 &  -17  $\pm$ 3 \\

C III 977 &   \leavevmode\change{169}  $\pm$     18  & \leavevmode\change{$>$ 13.6}  &  
 & [-80,45] \\

C III 977 &  &  14.43 $\pm$    0.47  
& 14   $\pm$ 4 & -15  $\pm$ 2 \\

%C II 903.96 &   78 $\pm$   24  & 13.65   $\pm$     0.04  &  
%[-80,45] & & \\

N II 1083 &  $<$ \leavevmode\change{96} & $<$ \leavevmode\change{13.9} &  
 & [-80,45] \\

N III 989 &  \leavevmode\change{65} $\pm$   20   & \leavevmode\change{13.96} $\pm$      \change{0.13}  &  
 & [-80,45] \\

N III 989  &  &  13.95 $\pm$    0.11  
& 14 $\pm$ 4  & -44  $\pm$ 3 \\

N IV 765 &  59 $\pm$ 17 & 13.32 $\pm$ 0.15    &  
 & [-80,45] \\
 
N IV 765 &  &  13.15 $\pm$ 0.18    
& 14  $\pm$ 4  & -12  $\pm$ 7\\

O II 834 &  65 $\pm$    12   &    13.97  $\pm$     0.08 &  
 & [-80,45]  \\

O II 834  &  &  13.97 $\pm$ 0.08    
& 14  $\pm$ 4  & -15  $\pm$ 3\\

O VI 1031 &  $<$  \leavevmode\change{57}  & $<$  13.6     &  
 & [-80,45]  \\

O VI 1037 &  $<$   \leavevmode\change{61}  & $<$  \leavevmode\change{13.9} &  
 & [-80,45]  \\

Ne VIII 770 &  $<$   \leavevmode\change{43} & $<$  \change{13.9}     &  
 & [-80,45]  \\

Ne VIII 780 &  $<$   \leavevmode\change{41} & $<$ \change{14.2} &  
 & [-80,45]  \\

Si II 1020 &  $<$   \leavevmode\change{58} & $<$ 14.6 &  
 & [-80,45]  \\

S IV 1062 &  $<$   \leavevmode\change{115} & $<$ 14.3      &  
 & [-80,45] \\

S V 786 &  $<$ \leavevmode\change{39} & $<$ 12.7    &  
 & [-80,45]  \\

S VI 933 &  $<$  \leavevmode\change{70} & $<$  13.3     &  
 & [-80,45]  \\
 
\bottomrule
\end{tabularx}
\end{table}

% Non detections: < (AOD measurement + error)

%$^{1}$ Smaller velocity interval has been considered because of close vicinity to Lyman series lines.

%$^{2}$ Feature is contaminated at v < -38 km/s. The $W_{r}$ and $N_{a}$ ranges are arrived at by integrating the apparent column density over the velocity intervals [-70,55] and [-38,55].

%$^{3}$ Velocity range slightly truncated to avoid contamination. Upper limit because it is a contaminating line.

%$^{4}$ Smaller velocity range to avoid contamination.Upper limit because it is a contaminating line.

%O VI 1037 has contamination;value used as upper limit

% B parameter of all lines are tied up to C II 1036

%%%%%%%%%%%%%%%%%%%%%%%%%%%%%%%%%%%%%%%%%%%%%%%%%%

%%%%%%%%%%%%%%%%%%%%%%%%%%%%%%%%%%%%%%%%%%%%%%%%%%

% Don't change these lines
\bsp	% typesetting comment
\label{lastpage}
\end{document}